\documentclass[12pt,graphicx]{article}
\usepackage{graphicx}
\usepackage{amsfonts}
\usepackage{amssymb}
\usepackage{dsfont}
\newcommand{\beqn}{\begin{eqnarray}}
\newcommand{\eeqn}{\end{eqnarray}}
\newcommand{\be}{\begin{equation}}
\newcommand{\ee}{\end{equation}}

\newcommand{\te}{\theta}
\newcommand{\thb}{\bar\theta}

\def\st{Stueckelberg~}
\def\s1{$s_{\alpha}$}
\def\s2{$s_{\gamma}$}
\def\s3{$s_{\delta}$}
\def\c1{$c_{\alpha}$}
\def\c2{$c_{\gamma}$}
\def\c3{$c_{\delta}$}

\def\y{Y_{\phi}}

\begin{document}

\thispagestyle{empty}

\begin{flushright}
\vspace{-3cm}
{\small CERN-PH-TH/2006-207 \\
        NUB-TH-3259\\
        v1 Oct 11 2006 \\[-.15cm]
}
\end{flushright}
\vspace{1cm}

\begin{center}

{\Large\bf{
 Extra-weakly Interacting Dark Matter }}

\vspace{1cm} {\bf Daniel Feldman\footnote{e-mail:
feldman.da@neu.edu}$^{,\dag}$}, {\bf Boris K\"ors}\footnote{e-mail:
kors@cern.ch}$^{*}$,  {\bf and Pran Nath}\footnote{e-mail:
nath@lepton.neu.edu}$^{,\dag}$ \vspace{.5cm}

\hbox{
\parbox{7cm}{
\begin{center}
{\it
$^*$ Physics Department, Theory Division, CERN \\
1211 Geneva 23, Switzerland }

\end{center}
}
\parbox{8cm}{
\begin{center}

{\it
$^{\dag}$Department of Physics \\
Northeastern University \\
Boston, Massachusetts 02115, USA \\
}

\end{center}
} } \vspace{2cm}

\end{center}

\begin{center}
{\bf Abstract} \\
\end{center}

We investigate a new type of dark matter with couplings to ordinary
matter naturally suppressed by at least 1 order of magnitude
compared to weak interactions. Despite the extra-weak interactions
massive particles of this type (XWIMPs) can satisfy the Wilkinson
Microwave Anisotropy Probe (WMAP) relic density constraints due to
coannihilation if their masses are close to that of the lightest
state of the minimal supersymmetric standard model (MSSM). The
region in the parameter space of a suitably extended minimal
supergravity (mSUGRA) model consistent with the WMAP3 constraints on
XWIMPs is determined. Plots for sparticles' masses are given which
can be subject to test at the Large Hadron Collider. As an example
for an explicit model we show that such a form of dark matter can
arise in certain $Z'$ extensions of the MSSM. Specifically we
consider an Abelian extension with spontaneous gauge symmetry
breaking via Fayet-Iliopoulos D-terms in the hidden sector. The LSP
of the full model arises from the extra $U(1)_X$ sector with
extra-weak couplings to Standard Model particles due to experimental
constraints. With R-parity conservation the new XWIMP is a candidate
for cold dark matter. In a certain limit the model reduces to the
Stueckelberg extension of the MSSM without a Higgs mechanism, and
wider ranges of models with similar characteristics are easy to
construct.
%


\clearpage \tableofcontents


\section{Introduction}

The nature of dark matter \cite{Clowe:2006eq} and dark energy
continues to be  one of the primary open questions in both particle
theory and cosmology. It is now widely believed that dark matter
must be constituted of particles outside the standard list of known
particles. Chief  among these are the so called  weakly interacting
massive  particles (WIMPS). Supersymmetry with R-parity conservation
leads naturally to such a particle in the form of the lightest
supersymmetric particle (LSP). In the framework of SUGRA unified
models the lightest neutralino is a particularly attractive
possibility.
\\

In this paper we  investigate  a new possibility in supersymmetric
models,  where the dark matter candidate  has extra-weak
interactions with matter;  interactions weaker than weak
interactions by at least one order of magnitude. We refer to these
particles as XWIMPs. We will show that such  a possibility can occur
naturally in certain extensions of the minimal supersymmetric
Standard Model (MSSM) by Abelian gauge symmetries $U(1)_X$. One such
class has been analyzed recently
\cite{kn1,kn2,kn3,kn3a,Feldman:2006ce,Feldman:2006wb}. Models of
this type are based on the \st mechanism which arises quite
naturally in string and D brane models. Further, some of the
specific features of  the models of
\cite{kn1,kn2,kn3,kn3a,Feldman:2006ce,Feldman:2006wb} and of the
type discussed here may have a string
realization\cite{Anastasopoulos:2006cz}. A more detailed discussion
of the motivation for such models may be found in the above
references.
 But  there
may be  a wider range of models where extra-weak dark matter can
appear. The relic density analysis of XWIMPS requires careful study
which will be discussed later in this paper.
\\

The basic elements of the models we discuss are exhibited in
Fig.(\ref{connector}) and involve three sectors: $(i)$ a visible
sector where the fields of the SM or the MSSM reside, $(ii)$ a
hidden sector which is neutral under the SM gauge group, and $(iii)$
a third sector \cite{Nath:1996qs} which is non-trivial under the SM
and the hidden sector gauge symmetry. Aside from gravity, the fields
of the visible sector and the hidden sector communicate only via
this  third sector which we therefore call the ``connector sector''.
Interactions with hidden particles can of course modify predictions
of the SM and are thus highly constrained by the precision data from
LEP and Tevatron, see e.g.\ \cite{Feldman:2006ce}.
\\

In the following, we  construct  explicit simple models where the
gauge group in the hidden sector is just an Abelian $U(1)_X$ with
spontaneous breaking and a massive $Z'$ gauge boson. Such Abelian
extensions of the MSSM occur in a wide class of models including
grand unified models, string and brane models
\cite{u1,ghi,Ibanez:2001nd,Anastasopoulos:2006cz,Dvali:1996rj,ad,kn4}.
The explicit elements of our first example are as follows:

\begin{enumerate}

\item
The visible sector contains gauge, matter and Higgs superfields of
the MSSM charged under
 the gauge group $SU(3)_C\times SU(2)_L\times U(1)_Y$, but neutral under
 $U(1)_X$.

\item
The hidden sector contains the gauge superfield for $U(1)_X$ which
are neutral under the Standard Model gauge group.

\item
The connector sector contains the chiral fields $\phi^{\pm}$ with
charges $\pm Q_X$ under $U(1)_X$ and $\pm Y_\phi$ under $U(1)_Y$.
They thus carry {\it{dual}} quantum numbers. The fields in the
visible
 and in the hidden sectors can communicate only via couplings with these connector fields.

\end{enumerate}

Spontaneous breaking of the $U(1)_X$ generates a mixing between the
hidden and  the visible fields. We will implement this breaking via
Fayet-Illiopoulos D-terms \cite{Fayet:1974jb}. The parameters that
measure the mixing are highly suppressed because of the precision
constraints on the electroweak predictions. Their smallness is
responsible for the extra-weak interactions of the  hidden and the
connector fields with the fields in the MSSM.
\\

If the LSP of the $U(1)_X$ sector which we call the XLSP is lighter
than the LSP of
 the visible sector, it will be the LSP of the  whole  system.\footnote{The alternative  acronym possibility EWIMP has already been used
 for a different type of WIMP\cite{Hisano:2003ec}.}
With R-parity conservation, it is then an XWIMP candidate for cold
dark matter.

We  will also show that the above class of models reduces in a
certain limit to the \st extension of MSSM introduced in \cite{kn2}.
It is also interesting to investigate if an XWIMP can arise in
models where mixing between the visible and the hidden sector occurs
in the gauge kinetic energy \cite{Holdom:1985ag}. In a
supersymmetrization of such a model with off-diagonal kinetic terms
one as well finds a mixing between the neutral fermions:  i.e.,  the
gauginos and the chiral  fermions of the visible and the hidden
sector. Thus, these models provide another class with potential of
an XWIMP responsible for dark matter. More generally, there may be a
much wider range of models with similar properties.
\\

The outline of the rest of the paper is as follows: In Sec.2 we work
out a $U(1)_X$ extension of the MSSM with symmetry breaking via
Fayet-Illiopoulos D-terms \cite{Fayet:1974jb}  mixing between the
$U(1)_X$ and the $U(1)_Y$ fields. The neutralino sector of this
system has a $6\times 6$ mass matrix. It can lead to an LSP with
extra-weak interactions composed mostly of neutral fermions in the
hidden sector. The model reduces  in a certain limit to the \st
extension of MSSM (the StMSSM) \cite{kn2}. We briefly discuss the
electroweak constraints on  the parameters  of the model. As an
alternative we next consider a mixing between the visible and hidden
sectors originating from the gauge kinetic energy which works very
similarly. In Sec. 3 we analyze the relic density of XWIMPs and show
that it is possible to satisfy the experimental constraints via the
process of coannihilation.  A detailed numerical  analysis  shows
that XWIMPs are candidates for cold dark matter consistent  with the
recent WMAP data. The sensitivity of the analysis on the errors in
the top quark mass under the constraints of radiative breaking of
the electroweak symmetry is also discussed.  Conclusions are given
in Sec. 4.


\section{Extra-weak dark matter in $Z'$ models}

To start with,  we introduce a class of extensions of the MSSM where
a natural  mixing of the neutral MSSM fields with fields from the
hidden sector appears via off-diagonal mass matrices. Towards the
end of this section we also discuss other possibilities to
facilitate a mixing of   a   very similar type.

\subsection{Broken $U(1)_X$ with Fayet-Iliopoulos terms}

A $U(1)_X$  extension of the MSSM with a Fayet-Iliopoulos (FI)
D-term can lead in a natural manner to extra-weakly interacting dark
matter constrained by LEP and Tevatron data. Specific $U(1)$
extensions have been studied quite extensively in the literature
\cite{Dvali:1996rj,u1,zpp}. The features of our model were already
explained in the introduction. The full gauge symmetry of the model
is $SU(3)_C\times SU(2)_L\times U(1)_Y\times U(1)_X$. It differs
from previous formulations in that a FI D-term breaks the extra
$U(1)$ gauge symmetry instead of an F-term. The Abelian vector
fields consist of the $U(1)_Y$ vector multiplet $(B_{\mu},
\lambda_B, D_B)$ and the $U(1)_X$ vector multiplet $(C_{\mu},
\lambda_C, D_C)$ with gauge kinetic Lagrangian given by\footnote{We
only use global supersymmetry here, not supergravity, and write
everything in Wess-Zumino gauge.}
\beqn {\cal L}_{\rm gkin} = -\frac{1}{4} B_{\mu\nu} B^{\mu\nu} - i
\lambda_B \sigma^{\mu}\partial_{\mu} \bar\lambda_B +\frac{1}{2}
D_B^2
-\frac{1}{4} C_{\mu\nu} C^{\mu\nu} - i \lambda_C
\sigma^{\mu}\partial_{\mu} \bar\lambda_C +\frac{1}{2} D_C^2 \ .
\eeqn
The superfields $\Phi^{\pm}$ with components $(\phi^{\pm}, f^{\pm},
F^{\pm})$ are described by
\beqn \label{matt} {\cal L}_{\Phi}&=& - |D_\mu \phi^+|^2 - i f^+
\sigma^\mu D_\mu \bar f^+ - \sqrt{2} \left(
 ig_X Q_X \phi^+ \bar f^+ \bar \lambda_C
 +ig_Y \y \phi^+ \bar f^+ \bar \lambda_B
                   +\, {\rm h.c.}\, \right)
\nonumber\\
&& +g_X D_C (\bar \phi^+ Q_X \phi^+) +  g_Y D_B (\bar \phi^+ \y
\phi^+)
+ \{ \Phi^+ \rightarrow \Phi^- \}\ , \eeqn
where $F^\pm$ is set to zero and the covariant derivatives of the
scalars are
\beqn D_\mu\phi^{\pm} = (\partial_\mu \pm ig_X Q_X C_\mu \pm ig_Y \y
B_\mu)\phi^{\pm}\ . \eeqn
Next we add to the mix the FI terms
 \beqn {\cal L}_{\rm FI}= \xi_X D_C + \xi_Y D_B\  .
 \eeqn
Elimination of the D terms gives us the scalar potential\footnote{As
was pointed out in \cite{EFK} this kind of spectrum together with FI
couplings leads to anomalies in supergravity which necessitates
field-dependent FI terms. We will here ignore the issue and only
deal with global supersymmetry explicitly.}
\beqn V= \frac{g_X^2}{2} \Big(Q_X|\phi^+|^2 -Q_X|\phi^-|^2
+\xi_X\Big)^2 +\frac{g_Y^2}{2} \Big(\y |\phi^+|^2 -\y |\phi^-|^2
+\xi_Y\Big)^2\  . \eeqn
Minimization of the potential leads to
\beqn
   \langle\phi^+\rangle=0\ , \quad
\langle\phi^-\rangle=\sqrt{\frac{g_X^2 \xi_X Q_X + g_Y^2 \xi_Y \y}
{g_X^2Q_X^2 + g_Y^2 \y^2}}\ .
   \label{phipm}
\eeqn We consider the bosonic sector first. Spontaneous breaking of
the electroweak symmetry gives rise to the mixing of the neutral
gauge fields $ (C^{\mu }, B^{\mu }, A^{3\mu })$, with $A^{\mu a}$
$(a=1,2,3)$ for the $SU(2)_L$ gauge fields. In this basis the mass
matrix in the neutral sector is of the form \be \left[
\begin{array}{ccc}
    M_1^2             & M_1M_2                                & 0 \\
    M_1M_2         & M_2^2 + \frac{1}{4}v^2g_Y^2  & -\frac{1}{4}v^2g_2g_Y \\
      0                   & -\frac{1}{4}v^2g_2g_Y            & \frac{1}{4}v^2g_2^2
\end{array}
\right]\ . \label{vmatrix} \ee The parameters $M_1$, $M_2$ are
defined so that
 \beqn
M_1= \sqrt 2 g_X Q_X \langle\phi^-\rangle\ ,\quad M_2=\sqrt 2 g_Y \y
\langle\phi^-\rangle \ . \eeqn There is a single massless mode, the
photon, and two massive modes the $Z$ and $Z'$.
  Since
$\langle\phi^+\rangle=0$, the superfield $\Phi^+$ does not enter in
the mixings in the mass  matrix for the fields in the hidden sector
and the fields in the visible sector, and we  do not consider it
further.
 The CP-even component of the complex scalar $\phi^-$ mixes with the two
CP-even Higgs fields of MSSM producing a $3\times 3$ mass matrix
similar to the analysis given in Ref. \cite{kn2,kn3}.
\\

In the neutral fermionic sector there are two additional mass
eigenstates beyond the four neutral fermionic states in the MSSM,
$\lambda_Y,\lambda_3, \tilde h_1, \tilde h_2$. One can reorganize
the Weyl spinors in terms of four-component Majorana spinors
$\chi_S$ (out of $f^-$) and $\lambda_X$ (out of $\lambda_C$) in a
standard way.
The $6\times 6$  neutralino mass matrix in the basis  $((\chi_S,
\lambda_X);(\lambda_Y, \lambda_3, \tilde h_1, \tilde h_2))$ reads
\beqn \label{neutrmass} \left[\matrix{ 0 & M_1 & M_2 & 0 & 0 & 0\cr
M_1& \tilde m_X & 0 & 0 & 0 & 0\cr M_2& 0 & \tilde m_1 & 0 &
-c_{\beta}s_{W}M_0 & s_{\beta}s_WM_0\cr 0 & 0 & 0 & \tilde m_2 &
c_{\beta}c_{W}M_0 & -s_{\beta}c_WM_0 \cr 0 & 0 & -c_{\beta}s_{W}M_0
&  c_{\beta}c_{W}M_0 & 0 & -\mu \cr 0 & 0 & s_{\beta}s_{W}M_0  &
-s_{\beta}c_{W}M_0 &  -\mu & 0}\right] \ . \label{neutralino} \eeqn
A few explanations are in order: $\tilde m_X$  arises from the soft
mass term $-\frac{1}{2} \tilde m_X\bar \lambda_X \lambda_X$, $M_0$
is the $Z$ boson mass at the tree level, $c_W=\cos\theta_W$,
$s_W=\sin\theta_W$ with $\theta_W$ the weak angle, similarly
$c_{\beta}=\cos\beta$, $s_{\beta}=\sin\beta$, with $\tan\beta=
\langle H_2\rangle /\langle H_1\rangle$, and finally
 $\mu$ the Higgs mixing parameter of the MSSM.
The mass eigenstates of  the system are defined as the following six
Majorana states
\beqn ((\xi_1^0, \xi_2^0) ;( \chi_1^0,\chi_2^0,\chi_3^0,\chi_4^0))\
, \eeqn
where $\chi_a^0$ $(a=1,2,3,4)$ are essentially the four neutralino
states of the MSSM and the $\xi_{\alpha}^0$, $(\alpha =1,2)$ the two
additional states composed mostly of the new neutral fermions.
\\

We will discuss in a moment that the current electroweak data  puts
a  stringent bound  on $\epsilon= M_2/M_1$ such that $|\epsilon| \ll
1$ \cite{Feldman:2006ce}. In this limit the masses  of $\xi_1^0,
\xi_2^0$ are
\beqn m_{\xi_1^0}\simeq \sqrt{M_1^{2} +\frac{1}{4}\tilde m_X^{2}}
-\frac{1}{2} \tilde m_X,~~ m_{\xi_2^0}\simeq\sqrt{M_1^{2}
+\frac{1}{4}\tilde m_X^{2}} +\frac{1}{2} \tilde m_X\ . \eeqn
For the case when the lightest of the MSSM neutralinos $\chi^0\equiv
\chi_1^0$ is also lighter than $\xi^0\equiv \xi_1^0$ nothing much
changes compared to the pure MSSM. The LSP of the MSSM will still be
the LSP of the full system, and the dark matter candidate  will be
essentially the same as  in the MSSM with minor modifications.
However, a very different scenario emerges if  $\xi^0$ is lighter
than $\chi^0$ and becomes the LSP. The upper bound on $\epsilon$
translates to a suppression of the couplings of $\xi^0$ to MSSM
fields relative to the couplings of $\chi^0$ by a factor of
$\epsilon$. Roughly speaking one can treat $\xi^0$ as a standard LSP
$\chi^0$ but with couplings appropriately suppressed by at least an
order of magnitude. This is why we call $\xi^0$ extra-weakly
interacting, an XWIMP.

\subsection{\st reduction of $U(1)_X$ extension}

In a certain limit the model of the previous subsection reduces to
the \st extension of MSSM proposed in \cite{kn2,kn3}. To achieve
the
 reduction   we assume  as is conventional in the analysis
of MSSM that $\xi_Y$ is negligible. We consider now the limit
$\langle\phi^-\rangle\rightarrow \infty$, $g_X Q_X\rightarrow 0$,
and  $g_Y Y_{\phi}\rightarrow 0$, with $M_1$ and $M_2$ fixed. This
leads to
\beqn \frac12 |D_{\mu}\phi^-|^2= \frac{1}{2} (M_1 C_{\mu}+ M_2
B_{\mu}
 +\partial_{\mu} a)^2  + \frac{1}{2} (\partial_{\mu}\rho)^2\ .
\eeqn
where $\phi^-= \rho+ia$. The Lagrangian can be written
$\mathcal{L}_{\Phi} = \mathcal{L}_{St} + \mathcal{L}_{\Phi^{+}}$ ,
where $\Phi^+$ is now completely decoupled from the vector multiplet
and ${\cal L}_{\rm St}$ can be written
\beqn \label{stueck} {\cal L}_{\rm St} &=& - \frac{1}{2}(M_1C_{\mu}
+M_2 B_{\mu} +\partial_{\mu} a)^2
 - \frac{1}{2} (\partial_\mu \rho)^2
- i \chi \sigma^{\mu} \partial_{\mu}\bar \chi
\\
&&  
 +\rho(M_1D_C +M_2 D_B)
 +\big [ \chi (M_1 \lambda_C + M_2 \lambda_B)
 + {\rm h.c.} \big]\ .
\nonumber \label{ls} \eeqn
This arises from the following density in superfield notation
\beqn {\cal L}_{\rm St} = \int d^2\te d^2\thb\ (M_1C+M_2B+  S +\bar
S )^2 \ , \label{mass} \eeqn where $C$ and $B$ are gauge
supermultiplets and $S$ a chiral supermultiplet. See \cite{kn2,kn3}
for more details. With the above one  then has exactly the  \st
extension of the MSSM.

\subsection{Electroweak constraints on mixing parameters}

The $U(1)_X$ extension of MSSM of the type discussed in Secs.(2.1)
and (2.2) has two mass parameters, $M_1$ and $M_2$, which can affect
electroweak physics. However,  since the Standard Model is already
in excellent agreement with LEP and Tevatron data the above
extensions  are severely constrained.   New physics  can only be
accommodated within the corridor of error bars consistent with
precision measurements.  For  example, the SM prediction relating
the $W$ and $Z$ masses (in the on-shell scheme) is given  by
\cite{marciano} \beqn \frac{ M_W^2}{M_Z^2}+  \frac{\pi \alpha}{
\sqrt 2 G_F M_W^2 (1-\Delta r)} =1\ .
 \label{wmass}
\eeqn In the above $\alpha$ is the  fine structure constant at the
scale $Q^2=0$, $G_F$ the Fermi constant, $\Delta r$ the radiative
correction such that $\Delta r= 0.0363\pm 0.0019$ \cite{:2005em},
where the  error in $\Delta r$ comes from the error in the top quark
mass and in $\alpha(M^2_Z)$. Using the current value of the $W$ mass
$M_W=(80.425 \pm 0.034)$ GeV \cite{:2005em} one finds the central
value of $M_Z$ in excellent agreement with the current data,
$M_Z=(91.1876\pm 0.0021)$ GeV. But the error of theoretical
prediction is $\delta M_Z \sim 30$ MeV.   Using  techniques similar
to the ones used in constraining extra  dimensions
\cite{Nath:1999fs} one may equate this error corridor in $M_Z$ to
the shift in $M_Z$ due to its coupling to the Stueckelberg sector of
the extended model. This constrains $\epsilon=M_2/M_1$  to lie in
the range \cite{Feldman:2006ce} \beqn
 |\epsilon| \lesssim0.061  \sqrt{1-\frac{M_Z^2}{M_1^2}}\label{epsilon}\ .
\eeqn A more detailed analysis of all the relevant precision
electroweak parameters can be found in
\cite{Feldman:2006ce,Feldman:2006wb}. For our current purposes it is
sufficient to know that a mixing parameter $\epsilon$ in the range
$0.1-0.01$ or smaller is in principle imaginable. This sets the
suppression factor for the couplings of XWIMPs relative to WIMPs in
our models.


\subsection{Abelian extension with off-diagonal kinetic terms}

There is a well known example of an Abelian extension of the SM with
a mixing between the visible and the hidden sector arising from an
off-diagonal kinetic energy \cite{Holdom:1985ag}. The hidden sector
in this model is called the shadow sector, the extra gauge factor
denoted $U(1)_S$. Specifically we write for the action
${\cal{L}}={\mathcal{L}}_{\rm SM}+ \Delta {\mathcal{L}}$, where
\beqn \Delta {\mathcal{L}}= -\frac{1}{4} C^{\mu\nu} C_{\mu\nu}
-\frac{\delta}{2} B^{\mu\nu}C_{\mu\nu} - |D_{\mu}\phi|^2- V(\phi,
\phi_{\rm SM})\ . \label{h1} \eeqn Here $C^{\mu}$ is gauge field for
the $U(1)_S$, $\phi$ is the Higgs charged under $U(1)_S$ giving mass
to $C^{\mu}$, and $\phi_{\rm SM}$ is the Standard Model Higgs. The
kinetic energy of Eq.(\ref{h1}) can be diagonalized by the
transformation \be\left(
\begin{array}{c}
    B^{\mu} \\
    C^{\mu}   \end{array}
\right)        = \left(
\begin{array}{cc}
 1 ~~ -s_{\delta}    \\
0 ~~~ ~c_{\delta}
  \end{array}
\right) \left(
\begin{array}{c}
    B^{\mu '}\\
    C^{\mu '}
    \end{array}
\right)\ ,\label{bcmatrix} \ee where $c_{\delta} =1/(1-\delta^2
)^{1/2}$, $s_{\delta} =\delta/(1-\delta^2)^{1/2}$. As in the
analysis of Refs. \cite{kn1,kn3,Feldman:2006ce,Feldman:2006wb} the
mixing parameter $\delta$ is small \cite{Kumar:2006gm,Chang:2006fp}.
After spontaneous breaking this type of model  also leads to a
massless photon, and two massive vector boson modes.
\\

To supersymmetrize the model we write the Lagrangian for the
extended theory ${\mathcal{L}} ={\mathcal{L}}_{\rm MSSM} +\Delta
{\mathcal{L}}$. In the pure gauge sector of the theory one has \beqn
\Delta {\mathcal{L}}_{\rm gkin}&=& - \frac{1}{4} C^{\mu\nu}
C_{\mu\nu}  -i  \lambda_C \sigma^{\mu} \partial_{\mu} \bar \lambda_C
+\frac{1}{2} D_C^2\nonumber\\
&& - \frac{\delta}{2} C^{\mu\nu} B_{\mu\nu}  -i \delta( \lambda_C
\sigma^{\mu} \partial_{\mu} \bar \lambda_B + \lambda_B \sigma^{\mu}
\partial_{\mu} \bar \lambda_C)
+ \delta D_BD_C\ . \eeqn One can give a mass to the $C_\mu$ by a
Stueckelberg mechanism without mixing with the hypercharge as in the
analysis of Ref.\cite{kn2}. Thus we add a term
 \beqn \Delta
{\mathcal{L}}_{\rm St}= \int d\theta^2 d\bar\theta^2 (M C + S +\bar
S)^2\ , \label{stmass} \eeqn where $C$ is the gauge multiplet for
the extra $U(1)_S$ and $S$ a chiral superfield. Everything works
very much the same way as in the standard \st extension. After
spontaneous breaking of the electroweak symmetry the neutralino mass
matrix in the basis $((\psi_S, \lambda_X'); (\lambda_Y', \lambda_3,
\tilde h_1, \tilde h_2))$, obtained after rotating the Majorana
fermions by the use of (\ref{bcmatrix}), is \beqn \left[\matrix{
    0 & M c_{\delta} & 0 & 0 & 0 & 0\cr
   M c_{\delta}& \tilde m_X c^2_{\delta}+\tilde m_1 s^2_{\delta} & -\tilde m_1 s_{\delta}  & 0 & s_{\delta}c_{\beta}s_{W}M_0 & -s_{\delta}s_{\beta}s_WM_0\cr
 0 &-\tilde m_1 s_{\delta} & \tilde m_1 & 0 & -c_{\beta}s_{W}M_0 & s_{\beta}s_WM_0\cr
  0 & 0 & 0 & \tilde m_2 & c_{\beta}c_{W}M_0 & -s_{\beta}c_WM_0 \cr
  0 &s_{\delta}c_{\beta}s_{W}M_0& -c_{\beta}s_{W}M_0  &  c_{\beta}c_{W}M_0 & 0 & -\mu \cr
   0 &    -s_{\delta}s_{\beta}s_WM_0   & s_{\beta}s_{W}M_0  &  -s_{\beta}c_{W}M_0 &  -\mu & 0}\right]\ .
\label{neutralino1} \eeqn The structure of the neutralino mass
matrix in Eq.(\ref{neutralino1}) is significantly different from
that of Eq.(\ref{neutralino}). Similar to the analysis of Sec.(2.1),
in the limit $s_{\delta}\to 0$ the states    $\psi_S$ and
$\lambda_X'$ decouple
 from the rest of the neutralinos. As before we label these two $\xi_1^0, \xi_2^0$ with
masses   given by
\beqn m_{\xi_1^0}\simeq \sqrt{M^{2} +\frac{1}{4}\tilde m_X^{2}}
-\frac{1}{2} \tilde m_X\ ,\quad m_{\xi_2^0}\simeq\sqrt{M^{2}
+\frac{1}{4}\tilde m_X^{2}} +\frac{1}{2} \tilde m_X\ . \eeqn
Diagonalizing Eq.(\ref{neutralino1}) one obtains six mass
eigenstates
$((\xi^0_1,\xi^0_2);(\chi^0_1,\chi^0_2,\chi^0_4,\chi^0_4))$. The
situation is very similar to the models discussed in previous
subsections with off-diagonal mass matrix. Thus we can summarize
that the supersymmetrized model with kinetic energy mixing can also
lead  to an XWIMP that becomes the XLSP with extra-weak coupling to
the Standard Model.
\\

From now on we use a unified notation labeling the extra-weakly
interacting particle as an arbitrary XWIMP denoting  any class of
model. The small mixing parameter will be called $\epsilon$ in any
case and the analysis of relic density  given below applies to all
such models with XWIMPs.


\section{Dark matter from XWIMPs}

Since the interactions of  XWIMPs with matter are extra-weak   the
annihilation of XWIMPs in general is much less efficient in the
early universe. Thus it requires some care to ascertain if a
reduction of the primordial density is possible in sufficient
amounts to satisfy the current relic density constraints. However,
the condition of thermal equilibrium are still satisfied for XWIMPs
as long as their interactions are only suppressed by few orders of
magnitude, say one or two. This requires that interaction rate
$\Gamma$ is  greater than the expansion rate  of the universe,
$\Gamma \geq H$ with $H=T^2/M_{\rm Pl}$. For the system at hand,
consisting of weakly and extra-weakly interacting massive particles
(WIMPS and XWIMPs) the condition of thermal equilibrium is indeed
satisfied. The XWIMPs will only slightly earlier fall out of
equilibrium but both types of species will be produced thermally
after the Big Bang or after inflation. This is in contrast to models
where the couplings of dark matter candidates are only of
gravitational strength or suppressed in similar ways.
\\

While the annihilation of XWIMPs alone cannot be sufficient to
deplete their density efficiently such reductions may be possible
with coannihilation \cite{gs}.  In  general, coannihilation could
involve all the neutralinos as well as squarks and sleptons in
processes of the type
 \beqn
\chi_i^0+\chi_j^0~~\to~~ f\bar f,\, WW,\, ZZ,\, Wh,\, \cdots~ \ ,
\label{coann} \eeqn where $\chi^0_i=(\xi_{\alpha}^0,\chi_{a}^0)$.
Let us explain how this can potentially lead to sufficient
annihilation of XWIMPs.
\\

The analysis of relic density involves the total number density of
neutralinos $n=\sum_i n_i$ which is governed by the Boltzman
equation \beqn \frac{dn}{dt} = -3Hn -\sum_{ij} \langle
\sigma_{ij}v\rangle (n_in_j-n_i^{\rm eq}n_j^{\rm eq}) \ , \eeqn
where $\sigma_{ij}$ is the cross-section for annihilation of
particle species $i,j$, and $n_i^{\rm eq}$ the number density of
$\chi^0_i$ in thermal equilibrium. The approximation $n_i/n=n_i^{\rm
eq}/n^{\rm eq}$ gives the well known
 \beqn
\frac{dn}{dt} =-3nH -\langle \sigma_{\rm eff}\rangle(n^2-(n^{\rm
eq})^2)\ , \eeqn where \beqn \sigma_{\rm eff}=\sum_{i,j}
\sigma_{ij}\gamma_i\gamma_j\ , \eeqn the $\gamma_i$ are the Boltzman
suppression factors \beqn \gamma_i=\frac{n_i^{\rm eq}}{n^{\rm eq}} =
\frac{g_i(1+\Delta_i)^{3/2} e^{-\Delta_i x}} {\sum_j g_j
(1+\Delta_j)^{3/2}e^{-\Delta_j x}}\ . \eeqn Here $g_i$ are the
degrees of freedom of $\chi_i$, $x={m_1}/{T}$ with $T$ the photon
temperature and $\Delta_i =(m_i-m_1)/m_1$, $m_1$ defined as the mass
of the XWIMP which is the LSP. The freeze-out temperature is given
by \beqn x_f =ln\Bigg[x^{-1/2}_f \langle \sigma_{eff} v \rangle m_1
\sqrt{\frac{45}{8 \pi^6 N_f G_N}} \hspace{.1cm}\Bigg].  \eeqn  Now
$N_f$ is the number of degrees of freedom at freeze-out and $G_N$ is
Newton's constant. The relic abundance  of XWIMPs at current
temperatures is finally \beqn \Omega_{\xi^0} h^2 = \frac {1.07\times
10^9 \rm{GeV}^{-1}}{N_f^{\frac{1}{2}} M_{\rm Pl} }
\left[\int_{x_f}^{\infty} \langle \sigma_{\rm eff} v\rangle
\frac{dx}{x^2}\right]^{-1}\ . \eeqn Here $x_f={m_1}/{T_f}$, $T_f$ is
the freeze-out temperature, $M_{\rm Pl} =1.2\times 10^{19}$ GeV and
$h$ the present day value of the Hubble parameter in the units of
100 $\rm{km \cdot s}^{-1} \cdot\rm{Mpc}^{-1}$.


\subsection{Relic density analysis for XWIMPs}

After all these preliminaries let us come to the specific treatment
of XWIMPs. The naive expectation is that XWIMPs would not be able to
annihilate in sufficient numbers to satisfy the current relic
density constraints.  An exception to this expectation is the
situation of coannihilation \cite{gs} that can drastically change
the picture. It can contribute in a very significant way to the
annihilation process.  Let us consider the coannihilation of a XWIMP
$\xi^0$ and a WIMP $\chi^0$ via the following set of processes \beqn
\xi^0+\xi^0&\to& X\ , \nonumber\\
\xi^0+\chi^0 &\to& X'\ , \nonumber\\
\chi^0+\chi^0&\to& X''\ , \eeqn where $\{X\}$ etc denote the
Standard Model final states.
The effective cross section in this case is \beqn \sigma_{\rm eff}=
\sigma_{\chi^0\chi^0}     \frac{1}{(1+Q)^2}  (Q
+\frac{\sigma_{\xi^0\chi^0} }{\sigma_{\chi^0\chi^0} })^2\ , \eeqn
where \beqn Q= \frac{g_{\chi^0}}{g_{\xi^0}} (1+\Delta)^{\frac{3}{2}}
e^{-x_f\Delta}\ . \eeqn Here $g$  is the degeneracy for the
corresponding particle and   $\Delta
=(m_{\chi^0}-m_{\xi^0})/m_{\xi^0}$.  For the case at hand, the ratio
${\sigma_{\xi^0\chi^0} }/{ \sigma_{\chi^0\chi^0} } \sim {\cal
O}(\epsilon^2)\ll 1$.  Thus if the mass  gap between $\xi^0$ and
$\chi^0$ is large so that $x_f\Delta \gg 1$, then  $\sigma_{\rm
eff}$ is much  smaller than the typical WIMP cross-section and  one
cannot annihilate the XWIMPs in an efficient  manner to satisfy the
relic density constraints.
\\

If the mass gap between the XWIMP and WIMP is small and the XWIMP is
still lighter than the WIMP we have the case of coannihilation. Let
us look at a parameter choice with $Q\sim O(1)$ and $Q\gg
{\sigma_{\xi^0\chi^0} }/{\sigma_{\chi^0\chi^0} }$. We can write
$\sigma_{\rm eff}$  in the form \beqn \sigma_{\rm eff}\simeq
\sigma_{\chi^0\chi^0} \left(\frac{Q}{1+Q}\right)^2 \ .
\label{reliceff} \eeqn The result of Eq.(\ref{reliceff}) is easily
extended under the same approximations including coannihilations
involving additional MSSM channels. Now Eq.(\ref{reliceff}) is
modified so that $\sigma_{\chi^0\chi^0}$  is replaced by  $
\sigma_{\rm eff}(MSSM)$ and $Q$ is defined so that $Q=\sum_{i=2}^{N}
Q_i$ where $Q_i=(g_i/g_1) (1+\Delta_i)^{3/2}e^{-x_f \Delta_i}$. When
$Q\sim {\cal O}(1)$ the XWIMP relic density is just a modification
of the WIMP relic density modified only by the multiplicative factor
$(\frac{Q}{1+Q})^2$. It is then possible to satisfy the relic
density constraints much in the same way as  one does  for the LSP
of MSSM.
\\

Nevertheless, the couplings of $\xi^0$ with quarks and leptons are
suppressed by a factor of $\epsilon$. Thus cross-sections for the
direct detection of dark matter will be suppressed by powers of the
mixing parameter, making the direct detection of the extra-weak dark
matter more difficult. However, $\xi^0$  will do as well as $\chi^0$
for the seeding of the galaxies.


\subsection{WMAP constraints on XWIMPs}

Extensive  analyses of the relic density for  WIMPS in
mSUGRA\cite{msugra} and its extensions exist in the literature.
Recent works \cite{recentdark,d1,d2}  have  shown that the WIMP
density in the MSSM and in SUGRA models can lie within the range
consistent with the WMAP data. The  three year data gives for the
relic density\cite{Spergel:2006hy} \beqn \Omega_{\rm CDM}
h^2=0.1045^{+0.0072}_{-0.0095} \hspace{.5cm} (\rm WMAP3)\ .
\label{wmap3} \eeqn We will impose this constraint using a $1\sigma$
corridor.
\\

The specific framework  we consider is a Abelian extension of mSUGRA
with a $U(1)_X$. This means, in the MSSM we use the mSUGRA framework
with the minimal set of characteristic parameters for the soft
breaking, i.e.\ the universal scalar mass $m_0$, the universal
gaugino mass $m_{1/2}$, the universal trilinear coupling $A_0$,
$\tan\beta$ and sign($\mu$).  This is our extended mSUGRA model.
 Its parameter space
is subject to several constraints including electroweak symmetry
breaking by renormalization group effects, and LEP and Tevatron
conditions on the sparticle spectrum. The most severe of the
collider data are from LEP referring to the light chargino mass
$m_{\chi^{+}_1}$ that is expected to be greater than 103.5 GeV
\cite{Abbiendi:2003sc}. We also exhibit the bound on the Higgs mass
which would eliminate $m_h<114.4$ GeV \cite{Barate:2003sz,LEP-limit}
although this is strictly valid only for the Standard Model.
Another stringent constraint on the parameter space arises from the
experimental limits on the process  $b\to s\gamma$. The current
experimental average value for the $BR(b\rightarrow s \gamma$) as
given by the {\it Heavy Flavor Averaging Group} \cite{hfag,bsg} is
\begin{equation}
{\rm BR}(b\rightarrow s \gamma)= (355 \pm 24_{-10}^{+9}\pm 3)\times
10^{-6} \ . \label{bsg1}
\end{equation}
In our numerical analysis we  take a fairly wide error corridor and
apply the constraints
\begin{equation}
 2.65 \times 10^{-4} \leq {\rm BR}(b\rightarrow s \gamma) \leq 4.45\times
 10^{-4}\ .
\end{equation}
For the calculation of the relic density of XWIMPS we use {\tt {
micrOMEGAS 2.0}}\cite{micro}, and for the  RGE computations {\tt
Suspect 2.3} package   of Ref. \cite{Djouadi:2002ze}. As a check on
the results we employ {\sffamily DarkSUSY 4.1} \cite{DS} along with
the {\tt Isajet 7.69} package of Ref. \cite{Paige:2003mg} and {\tt
ISASUGRA 7.69}. In our scans of the parameter space we implement a
Monte Carlo technique for each scan, sampling up to $10^6$ points
per scan. As input GUT scale parameters we take $m_{1/2} \in
(0,1.5)$ {TeV}, $m_0 \in (0,3.5)$ {TeV}, $A_0 = 0$, and multiple
values of $\tan \beta \in (10-50)$, while sign$(\mu)$ is always
positive. Such regions of the mSUGRA
 parameter space are within the reach of the
LHC \cite{lhc,deBoer:2006bg} and the Tevatron \cite{tevatron}. The
particular  values of $\tan \beta$ favor such a scenario. In the
analysis of the relic density of XWIMPs we have chosen $\Delta$ to
lie in the range $(0,0.1)$.  Further, the contribution of the $Z'$
pole is included using the method of Ref. \cite{Nath:1992ty}.
However, its contribution turns out to be rather small due to the
small width of the $Z'$ and thus no appreciable effects  occur. For
the top mass the central value of the most recent evaluation from
the CDF and D0 collaborations is \cite{Group:2006xn}. \beqn m_t=
171.4 \pm 2.1 ~{\rm GeV}\ . \label{topmass} \eeqn The bottom mass is
taken to be  $m_b(m_b)=4.23$ GeV. The allowed parameter space of
mSUGRA is very sensitive to the assumed value of the top quark mass
and of the bottom quark mass, and thus the dark matter analyses are
also very sensitive to these inputs \cite{sensitivity}.  We will
discuss this issue in further detail at the end of this section.
\\

In the calculation of the relic density, we find in general good
agreement between {\sffamily DarkSUSY 4.1} and {\tt { micrOMEGAS
2.0}} (up to about 15\%)
 for values of $\tan \beta$ in the range $(10-40)$.
 The main result is that the WMAP3 constraints are satisfied
 by XWIMPs for a wide range of $\tan\beta$, even though the
 allowed parameter space consistent with all the constraints
 does depend on the value of $\tan\beta$.
 We consider two representative  values in this paper, namely $\tan\beta =30,50$.
\\

We now discuss the details of the analysis. In Fig.(\ref{fset1}) we
display the relic density constraints on the XWIMPs in the
$m_0-m_{1/2}$ plane for the cases  $\tan\beta =30,50$ consistent
with Eq.(\ref{wmap3}).
  The black region satisfies the relic density constraints which lie within
 $1\sigma$ corridor of the central value of Eq.(\ref{wmap3}), while
  the shaded regions are eliminated due to other constraints. The  other constraints
  arise mainly from the lower limit on the chargino mass and the
 $b\to s \gamma$ branching ratio. The bound on the  Higgs mass
is also shown but only a small   additional region
  of the parameter space is  eliminated.
The analysis of Fig.(\ref{fset1}) shows that the relic density is
satisfied in both a low $m_0$ region, where one has  typically
coannihilation between the lightest neutralino  of the MSSM and the
stau, and a high $m_0$ region, which is characteristically the
hyperbolic branch (HB) of  radiative breaking of the electroweak
symmetry \cite{hb}, where the LSP and the next to lowest
supersymmetric particle (NLSP) become degenerate and are mostly
higgsino like.
\\

In Fig.(\ref{fset2}) we show the parameter space in the
$m_{\chi^0}-m_{1/2}$ and the $m_{\chi^+}-m_{\chi^0}$ plane.
  These plots display  the regions where  scaling
holds or breaks down which are also good indicators of the  gaugino
vs.\ higgsino composition  of $\chi^0$ (the LSP of the MSSM). Thus
in the $m_{\chi^0}-m_{1/2}$ plot points on  the straight line
boundary satisfy the scaling phenomenon,
 where  $m_{\chi^0} \simeq 0.5  m_{1/2}$. Here   $m_{\chi^0}$ is mostly a Bino. More  generally,
  scaling \cite{scaling} gives $m_{\tilde g}:m_{\chi_1^{+}}: m_{\chi^0}\simeq (6-7): 2:1$.
 On the other hand, when $m_{\chi^0}/m_{1/2}$ is significantly smaller
 than $0.5$ $\chi^0$ has a large higgsino component and typically
 arises from the HB.
A similar situation arises in the  $m_{\chi^+}-m_{\chi^0}$ plane.
The points on the upper straight line satisfy scaling, while those
on the lower curved area have a large higgsino component and thus
violate scaling.
\\

In Fig.(\ref{fset3}) we exhibit the allowed parameter space in the
$m_{\tilde g}-m_{\chi^{+}}$ plane. On the lower straight line along
the diagonal $\chi^0$ is Bino-like and the scaling relation
$m_{\tilde g}: m_{\chi^0}= (6-7):1$ is satisfied. Above this region
$\chi^0$ has a significant higgsino component, and scaling is
violated. Further, in Fig.(\ref{fset3}) one can find the allowed
region in the $m_{\tilde t_1}-m_{\tilde g}$ plane and, finally, in
 Fig.(\ref{fset4}) the allowed region
in the $m_{\tilde t_1}-m_{\chi^+}$ and in the $m_h-m_{\chi^+}$ plot.
All figures show that the permissable mass range for the light stop
$\tilde t_1$ is rather wide, while for the Higgs there is a narrow
window. Typically, its mass has to lie within the corridor from the
lower limit of 114 GeV up  to about 125 GeV.


Let us add a comment regarding the impact of experimental error bars
on the top mass  under the constraints of the electroweak symmetry
breaking.
 As indicated above,  the region in the
parameter space of mSUGRA consistent with electroweak symmetry
breaking depends very sensitively on the mass of the top quark, a
phenomena which has been known for some time and which affects the
relic density \cite{sensitivity,d2}.
We emphasize that the sensitivity of the relic density arises because the sparticle
spectrum in SUGRA unified models, where the sparticle spectrum arises as a
consequence of radiative breaking of the electroweak symmetry (REWSB),
 is very sensitive to the top  mass.  This can be seen, for example, in the first paper of
  Ref.\cite{sensitivity} where it is shown that the stop mass can turn
tachyonic with variations in the top mass under constraints of  REWSB.
However, in MSSM scenarios where one can fix
the sparticle spectrum  and vary the top mass, the relic density is not
sensitive to variations in the top mass.  In contrast,  in the current analysis
the sensitivities to the top mass arise since we are using the framework of SUGRA
unification where the spectrum is computed via REWSB.
The recent  more accurate
determinations of the top mass have now very much reduced the error.
In the analysis of Figs.(\ref{fset1}-\ref{fset4}) we have used the
central value of Eq.(\ref{topmass}). We now consider a $1\sigma$
variation around this central value. Thus the results displayed in
Figs.(\ref{fset5}) are stated for  $m_t=169.3$ GeV, a $1\sigma$
downward
 shift on the central value, while those of
Figs.(\ref{fset6}) are for $m_t=173.5$ GeV which involve a $1\sigma$
upward
 shift.
   Quite remarkably one finds  that even a $1\sigma$ variation with reduced error bars  generates
very significant changes in the relic density. Specifically, a lower
top mass implies a larger portion in parameter space consistent with
the constraints.


\section{Conclusion}

We have introduced a new dark matter candidate whose interactions
with Standard Model matter are extra-weak, weaker than weak
interactions by at least one order of magnitude. Extra-weakly
interacting particles can arise in  a wide range of models, $Z'$
extensions of the MSSM with Higgs sectors, in the \st extension, in
extensions of the MSSM with off-diagonal gauge boson kinetic terms,
and possibly many other realization of small mixing between visible
and hidden sector fields. The new XWIMPs are good candidates for
dark matter if they become the LSP of the full system, in spite of
the extra-weak interactions with the MSSM. They can satisfy the
relic density constraints consistent with the WMAP data via
coannihilation. A direct observation of XWIMPs in dark matter
detectors will be more difficult. However, indirect tests of the
model are possible and should be investigated.

\vspace{1cm}
\begin{center}
{\bf Acknowledgments}
\end{center}
 We would like to thank Paolo Gondolo for communications regarding the most
  recent version of the  package  DarkSUSY and  Alexander Pukhov for  a communication regarding the
 most recent version of the  package micrOMEGAS.  We also thank  Mario Gomez
for useful discussions and communications regarding software
packages.
 We thank the  Opportunity Research Computing
Cluster of the Academic Research Computing User Group at
Northeastern University for the allocation of significant
supercomputer time for the numerical analysis of the relic density
in this work.  The work of D.~F. and P.~N. was supported in part by
the U.S. National Science Foundation under the grant
NSF-PHY-0546568.
\\

\clearpage
\section{Figures}
\vspace{.5cm}
 {\bf A PDF viewer is recommended\footnote{Higher
resolution eps figures of larger file sizes are available.}

\begin{figure}[h]
\vspace{1.0in} \hspace{-.2in} \centering
\includegraphics[width=10cm,height=8cm]{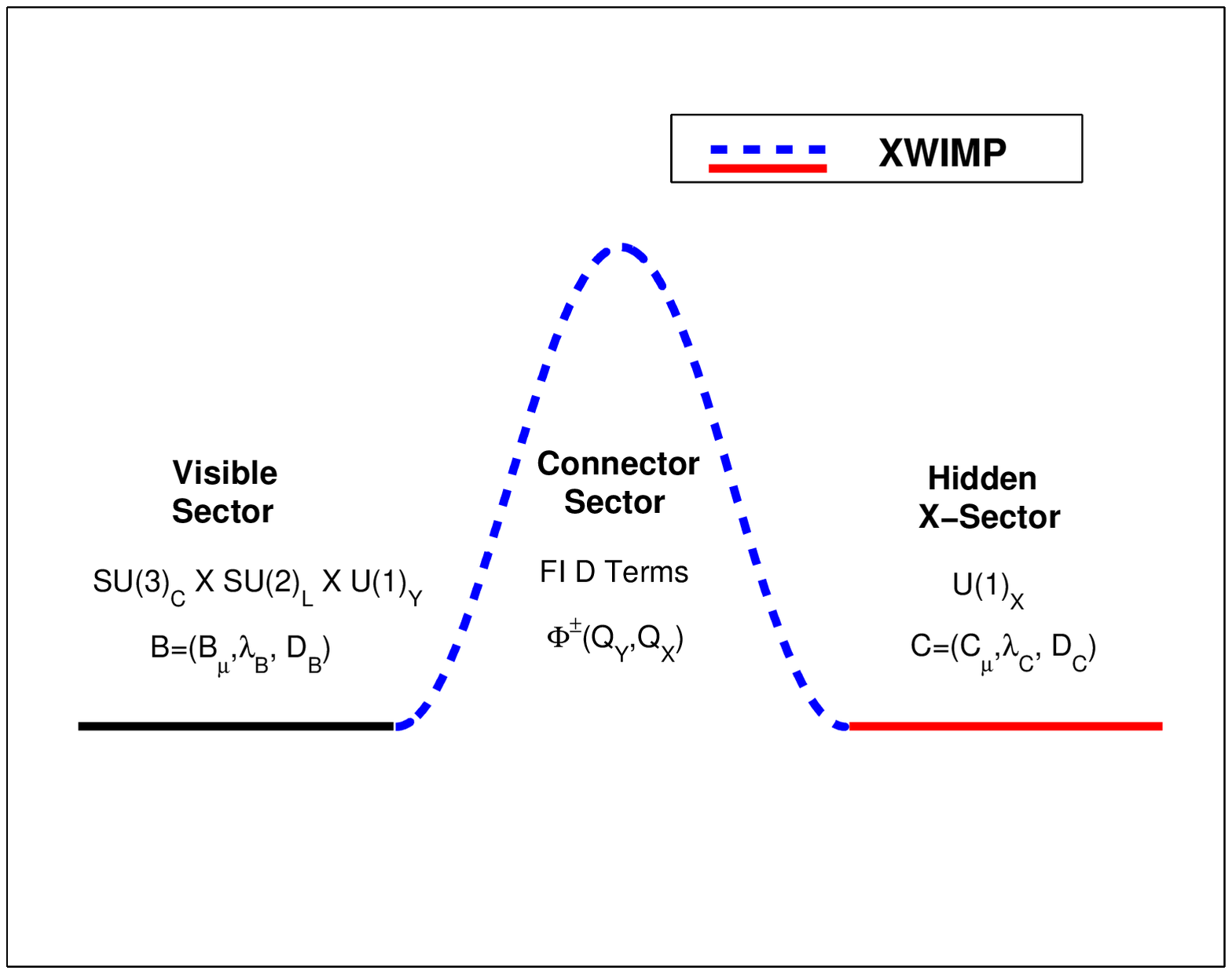}
\caption{The generation of the extra-weakly interacting massive
particles. An  XWIMP is a linear combination of fields in the hidden
sector and in the connector sector, and its interactions with the
MSSM particles are suppressed.}
 \label{connector}
\end{figure}

\clearpage

\begin{figure}[h]
\hspace*{-.2in} \centering
\includegraphics[width=12cm,height=9cm]{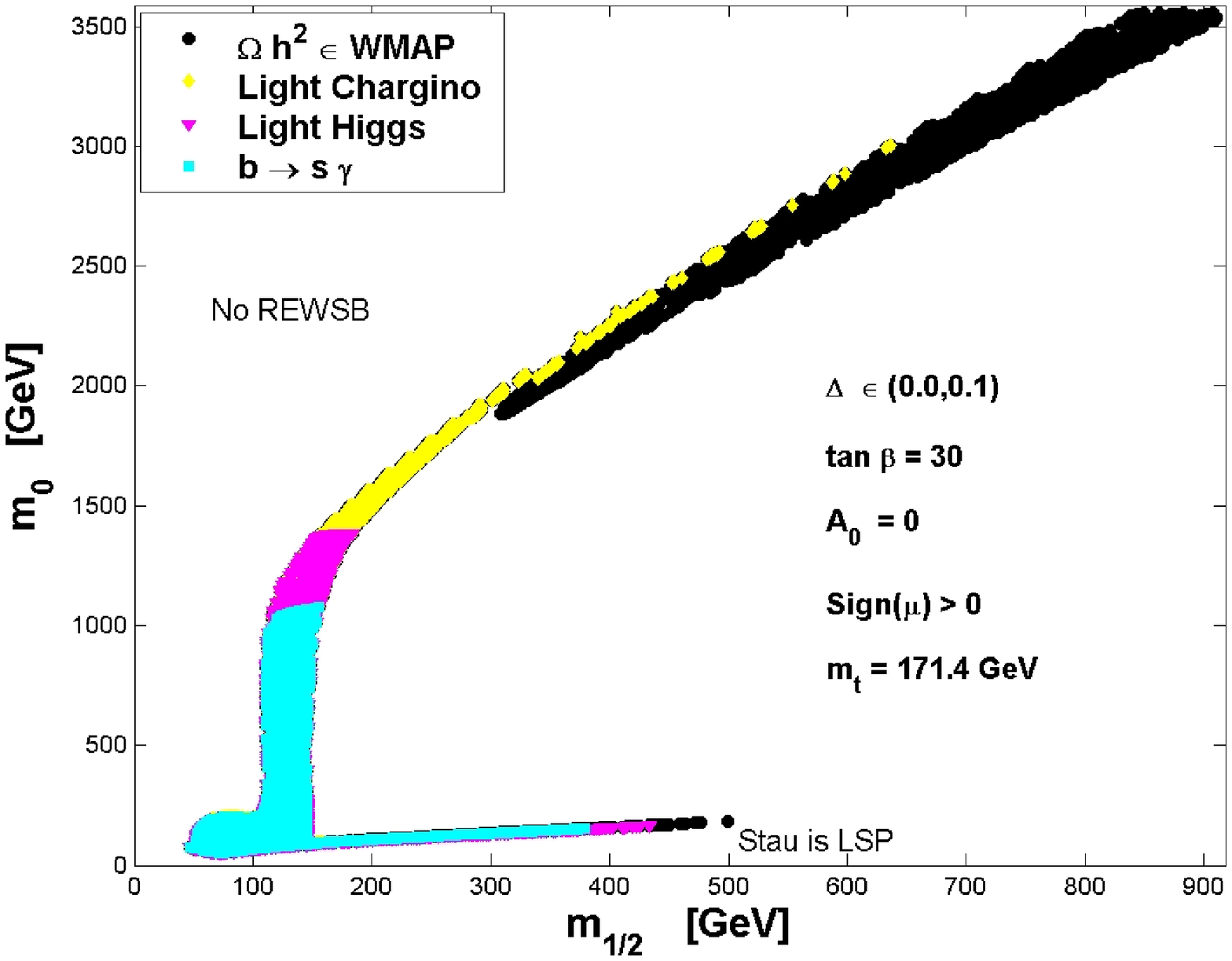}
\includegraphics[width=12cm,height=9cm]{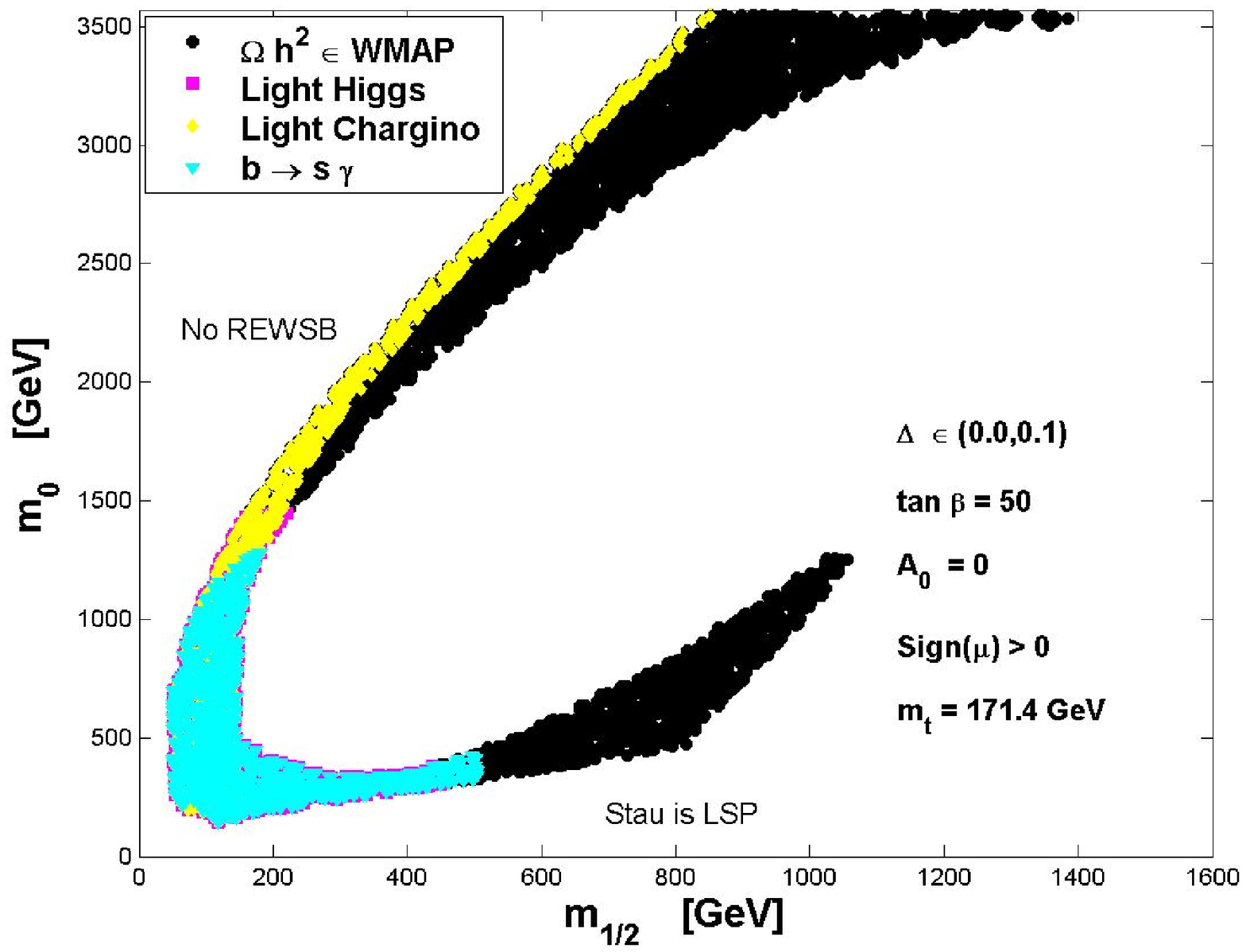}
\caption{The allowed parameter space in the  $m_0-m_{1/2}$ plane,
 under  the $1\sigma$ WMAP3 constraint of  Eq.(\ref{wmap3}) in extended mSUGRA
 for the case $A_0=0$, $\tan \beta = (30,50)$ (upper,lower),  sign($\mu>0$), $m_t=
171.4$ GeV,  $m_{1/2}\in (0,1.5)$TeV and $m_0 \in (0,3.5) $TeV, and
$\Delta$ in the range (0.0, 0.1).
  Regions eliminated  by the light chargino mass constraint, by the light
Higgs mass constraint, and by the $b\to s+\gamma$ constraint are
also exhibited.  } \label{fset1}
\end{figure}
%
%

\clearpage

\begin{figure}[h]
 \vspace*{1in}\hspace*{-.2in} \centering
\includegraphics[width=7.3cm,height=6.1cm]{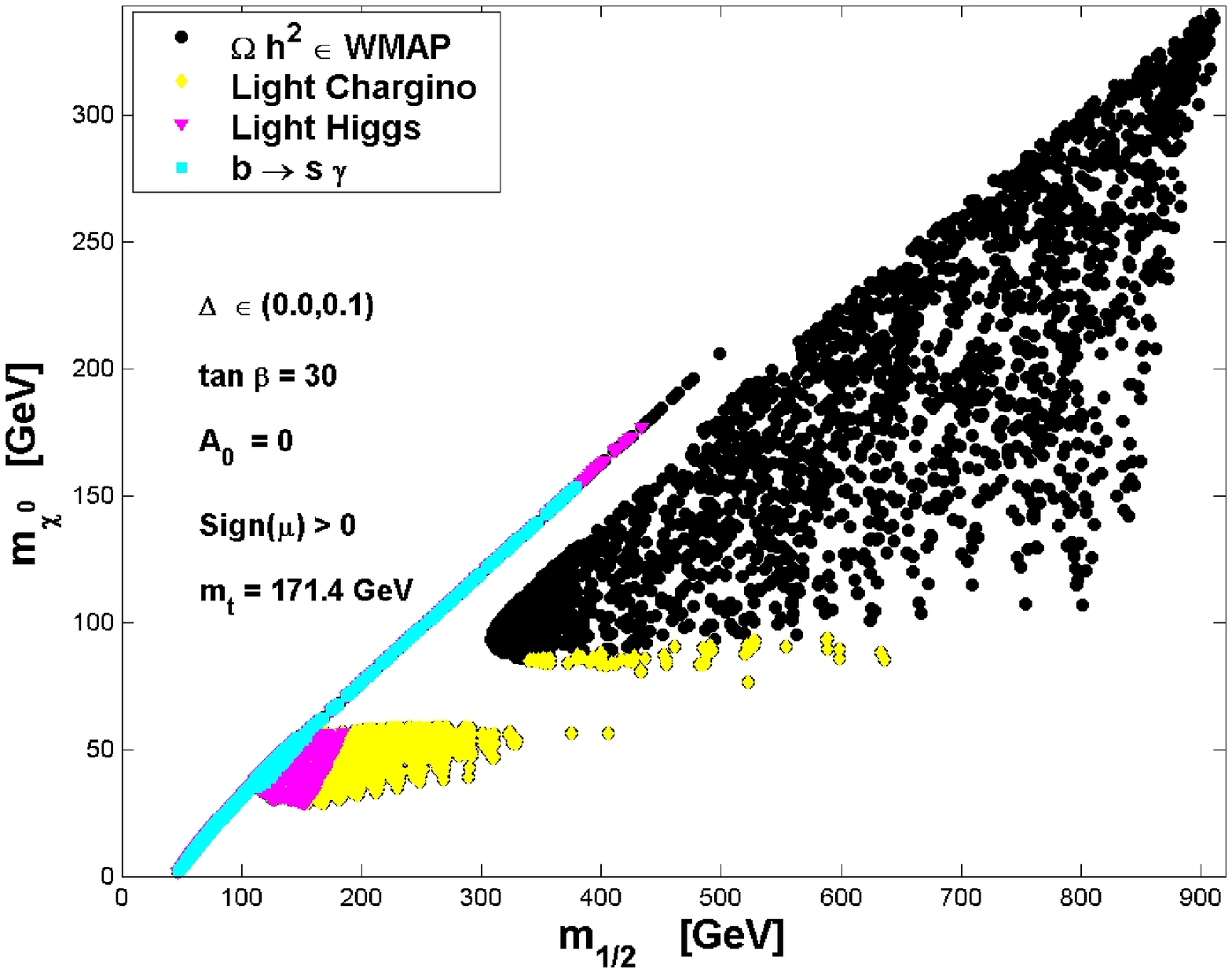}
\includegraphics[width=7.3cm,height=6.1cm]{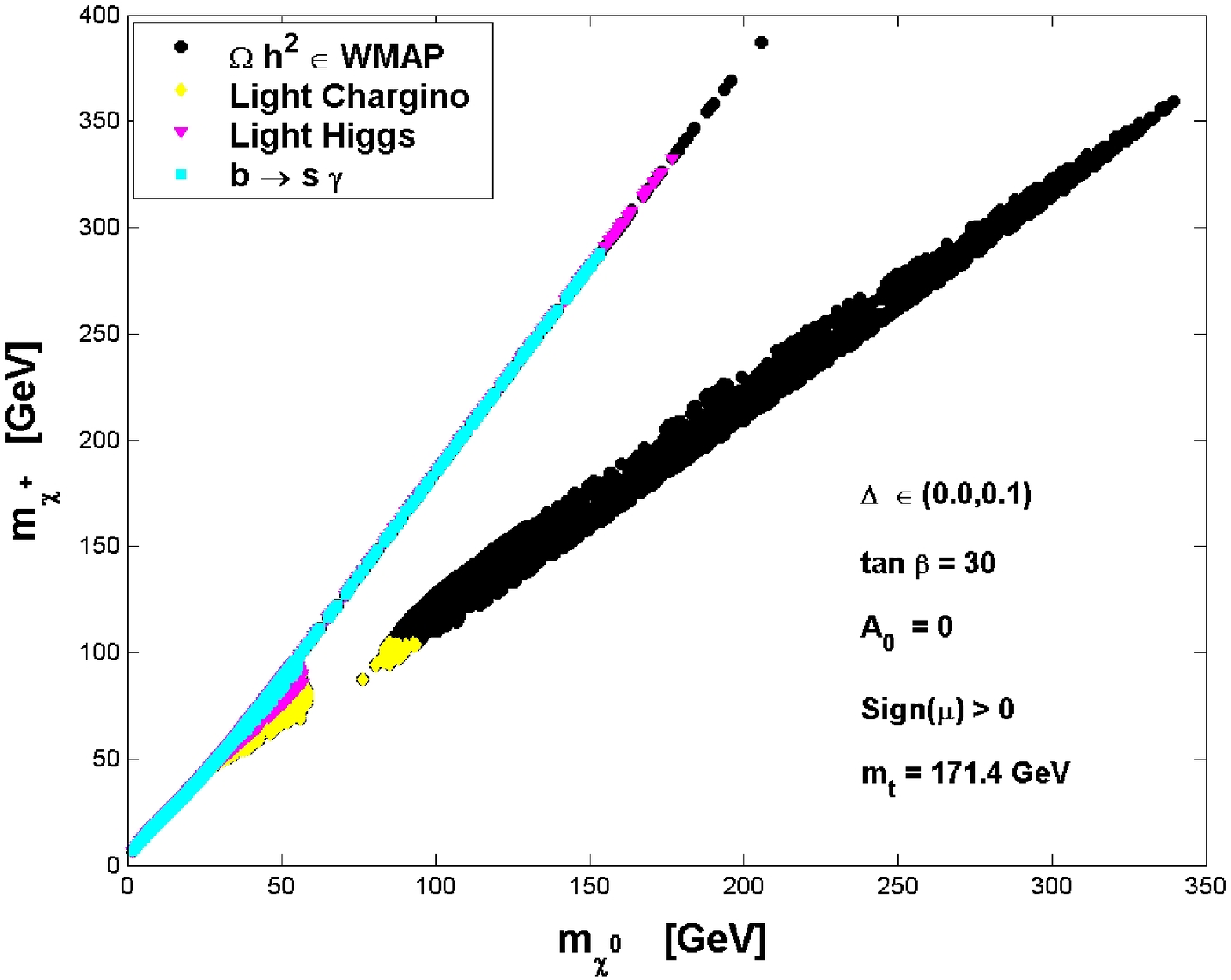}
\includegraphics[width=7.3cm,height=6.1cm]{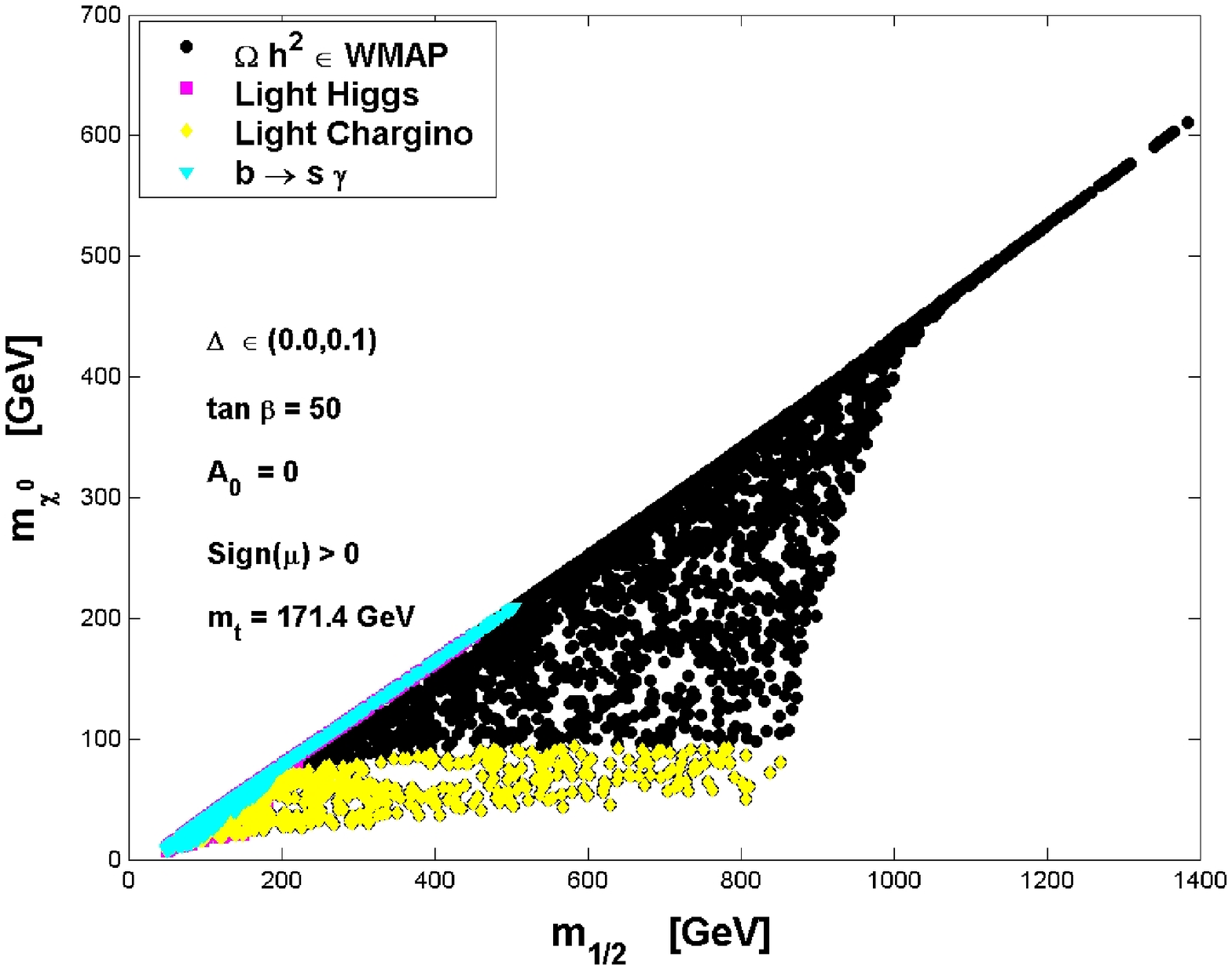}
\includegraphics[width=7.3cm,height=6.1cm]{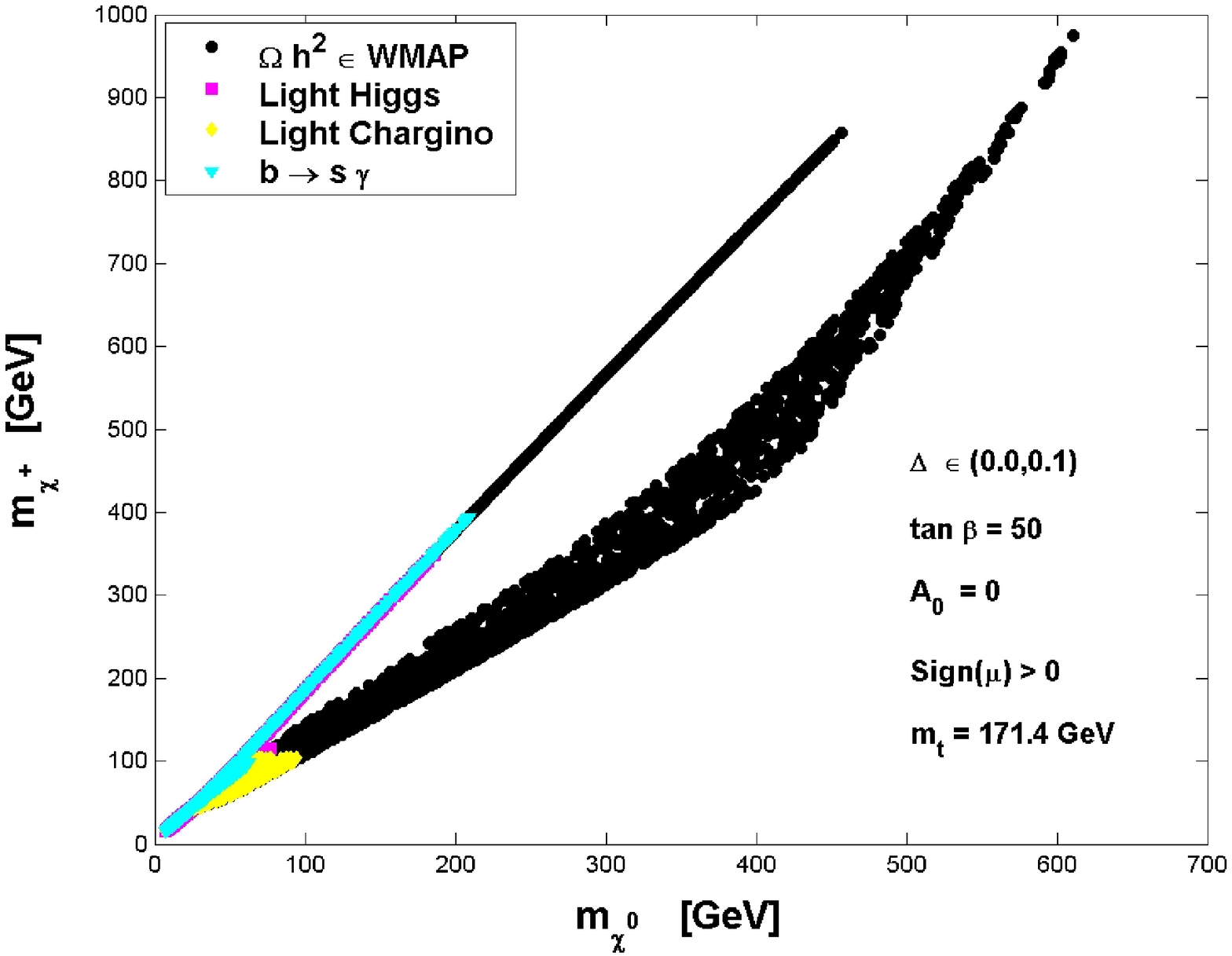}
\caption{The allowed parameter space in the  $m_{\chi^0}-m_{1/2}$
plane and in the  $m_{\chi^{+}}-m_{\chi^{0}}$   plane,
 under  the  $1\sigma$ WMAP3 constraint of  Eq.(\ref{wmap3}) in  extended mSUGRA
 for the same data set as  in Fig.(\ref{fset1}). These plots exhibit the phenomenon
of scaling and its  breakdown and in particular for the ratio
$m_{\chi^+}/ m_{\chi^0}$ as discussed in the text.}
 \label{fset2}
\end{figure}
%
%
\begin{figure}[b]
\vspace*{1in} \hspace*{-.2in} \centering
\includegraphics[width=7.3cm,height=6.1cm]{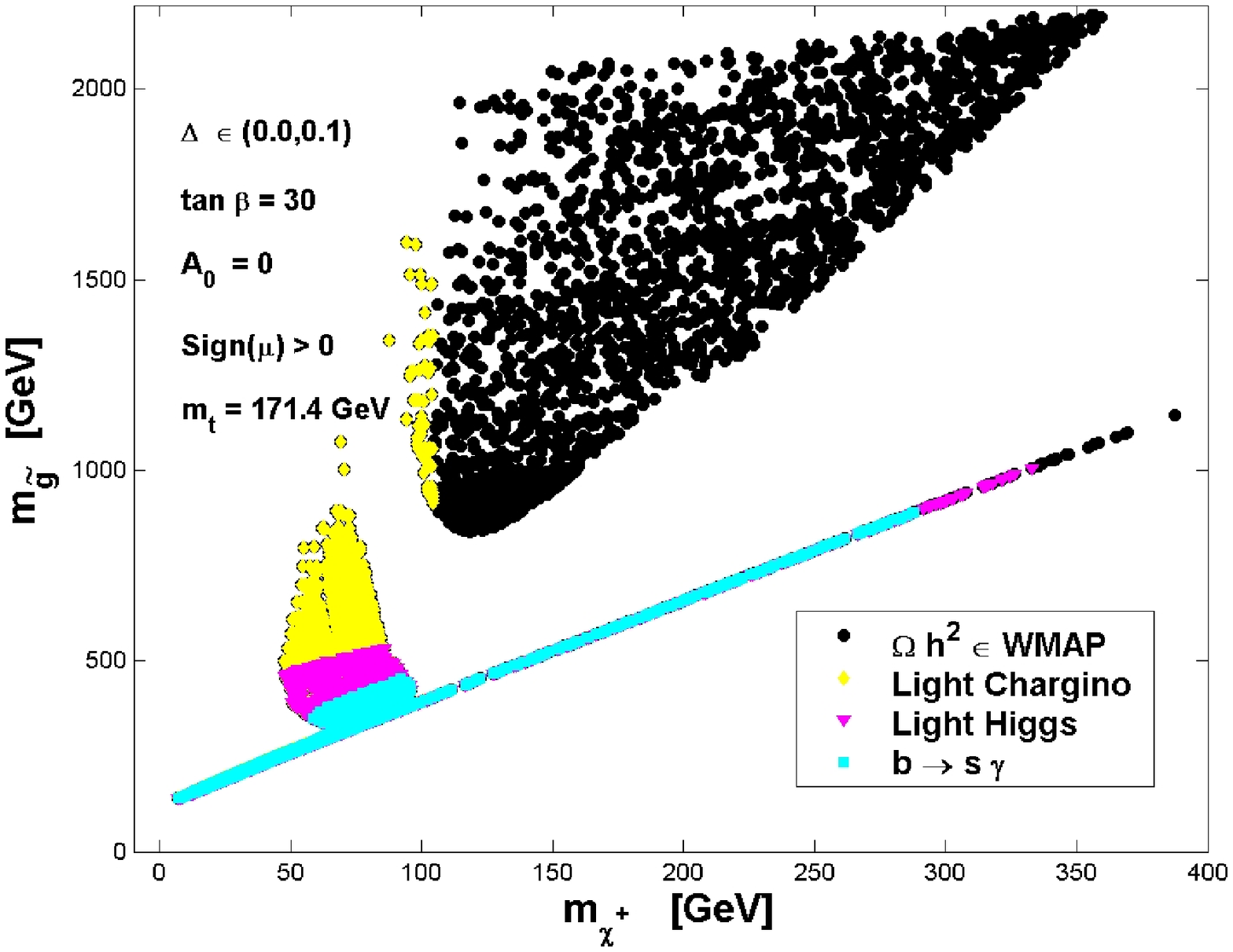}
\includegraphics[width=7.3cm,height=6.1cm]{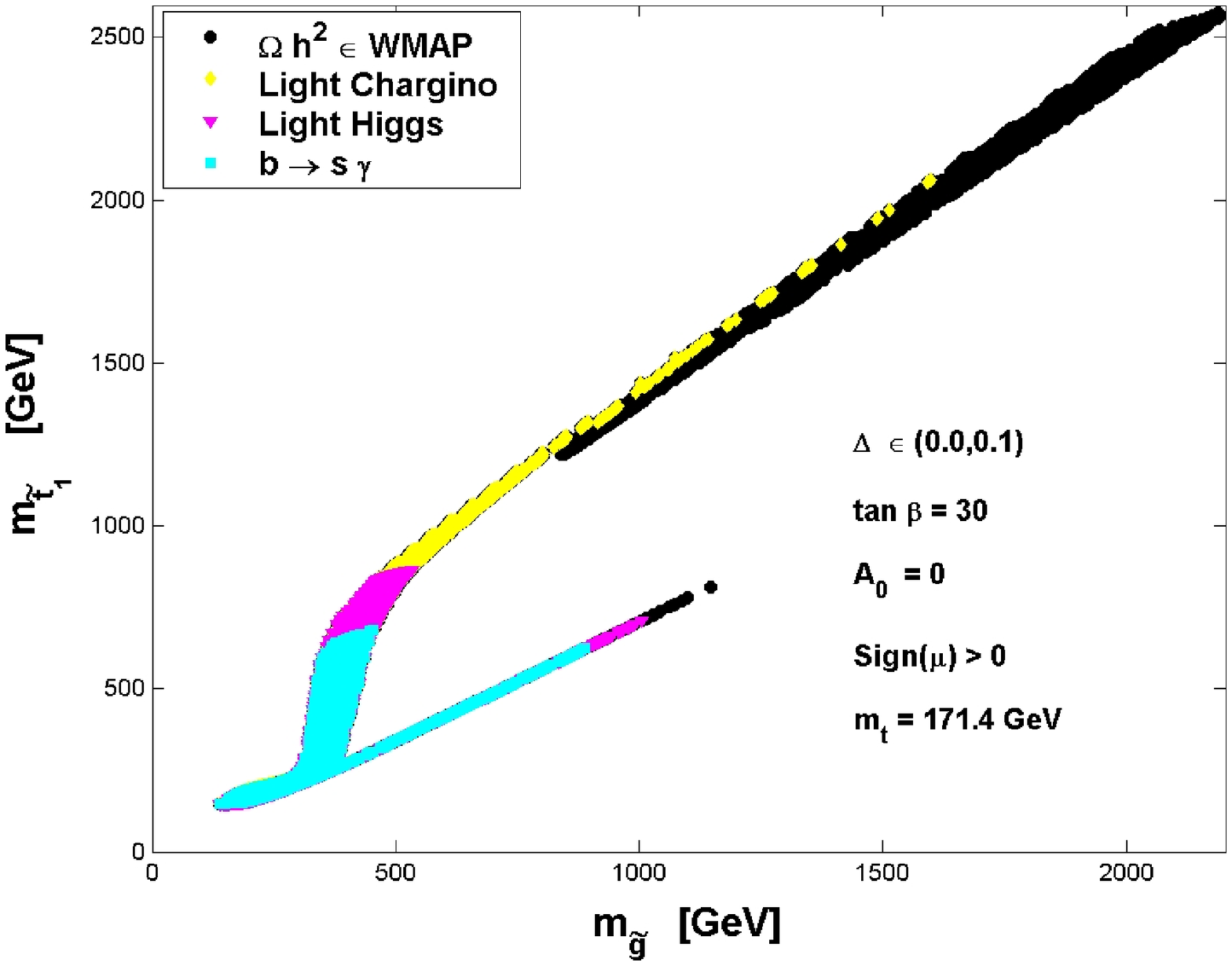}
\includegraphics[width=7.3cm,height=6.1cm]{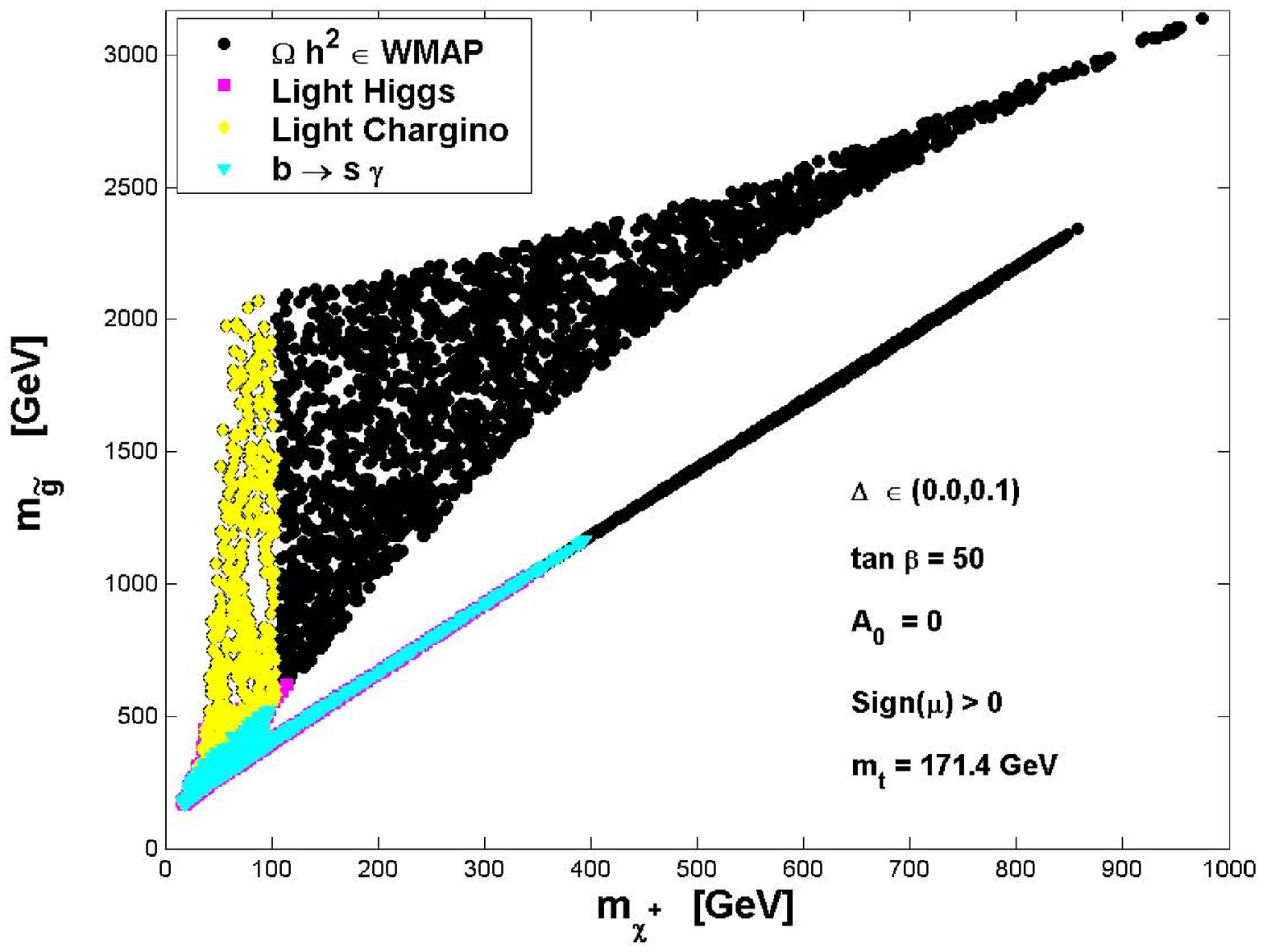}
\includegraphics[width=7.3cm,height=6.1cm]{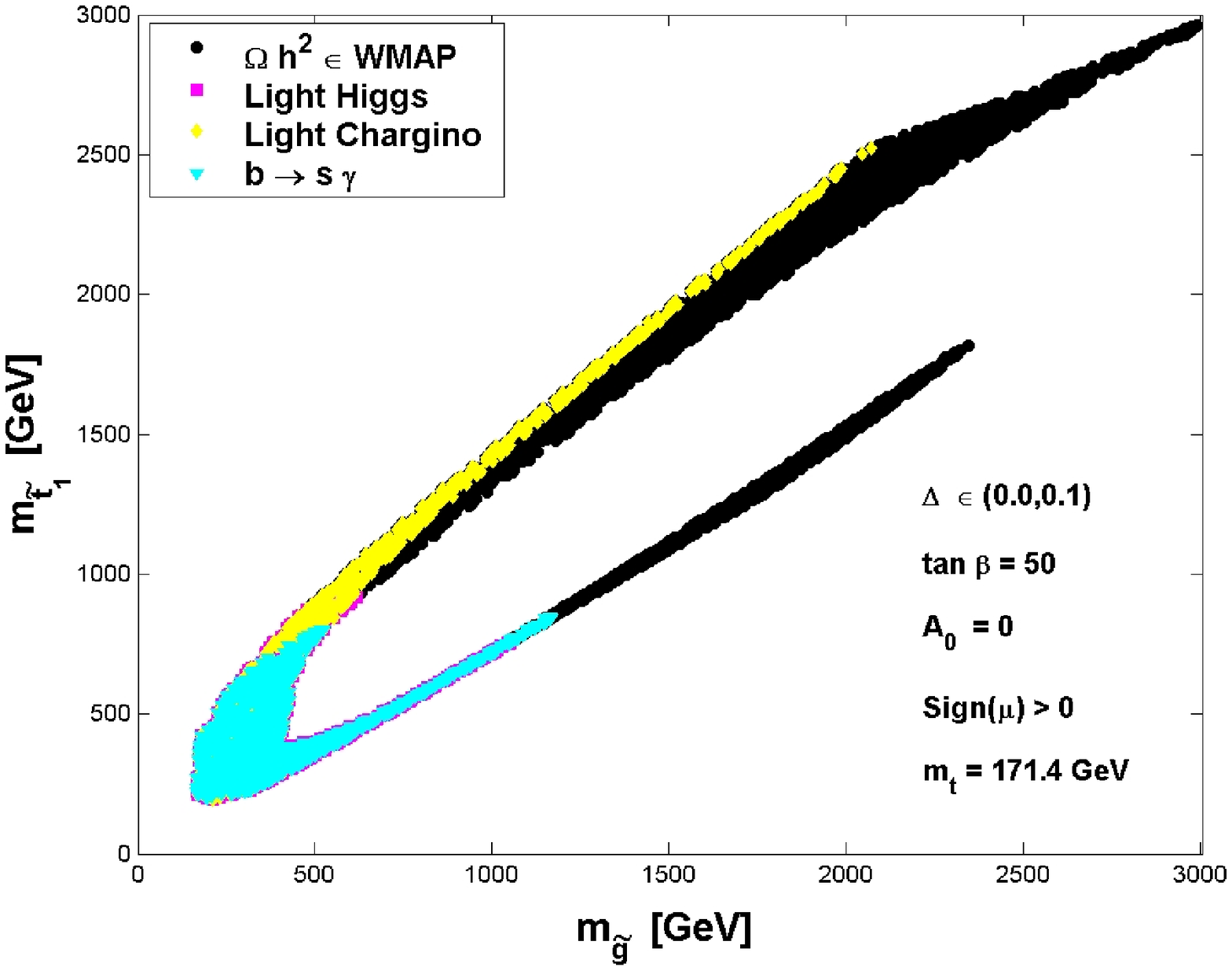}
\caption{The allowed parameter space in the  $m_{\tilde
g}-m_{\chi^+}$ plane and in the   $m_{\tilde t_1} -m_{\tilde g}$
plane,  under  the $1\sigma$ WMAP3 constraint of  Eq.(\ref{wmap3})
in  extended mSUGRA
 for the same data set as  in Fig.(\ref{fset1}).   }
 \label{fset3}
\end{figure}
\clearpage

\begin{figure}[b]
\vspace*{1in} \hspace*{-.2in} \centering
\includegraphics[width=7.3cm,height=6.1cm]{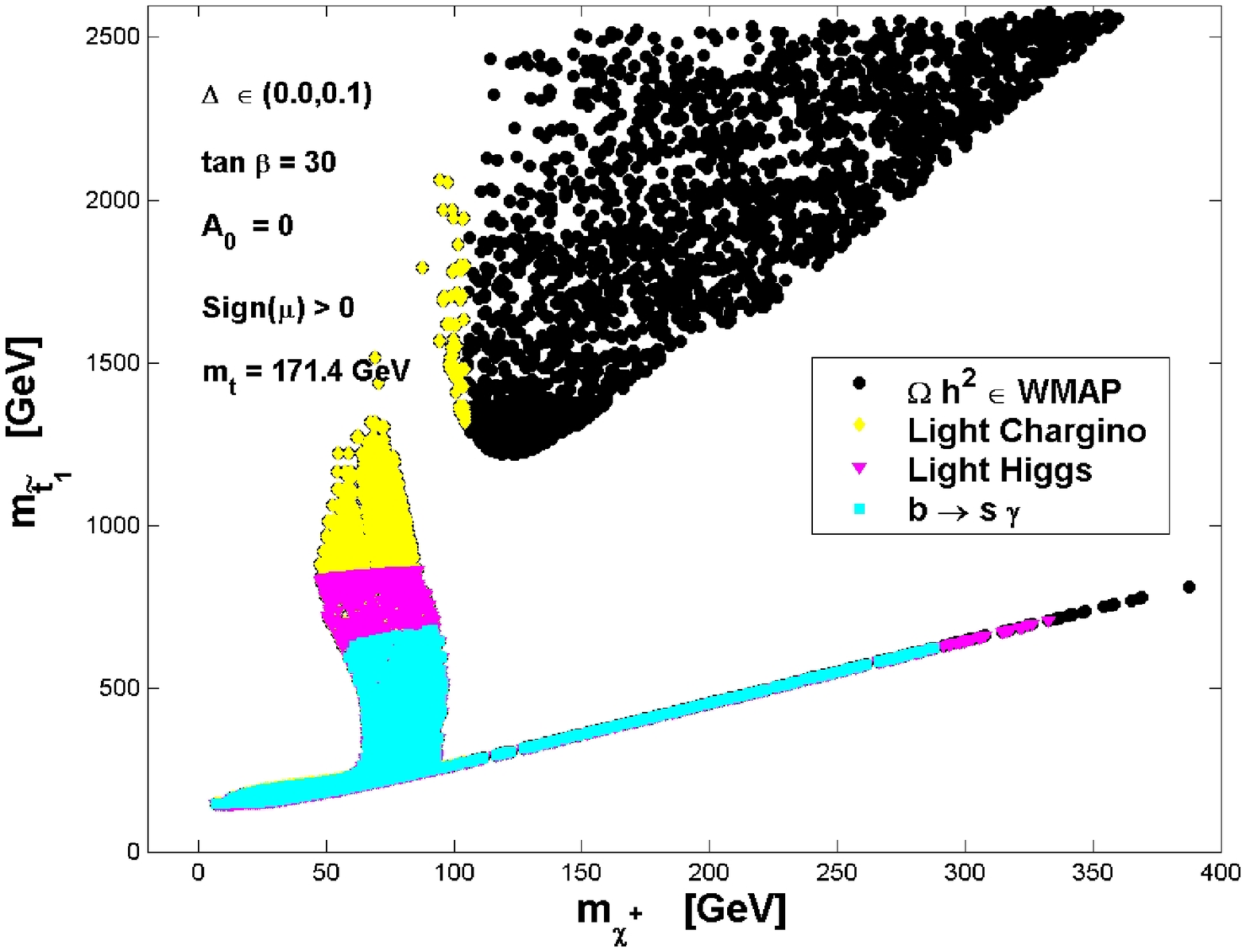}
\includegraphics[width=7.3cm,height=6.1cm]{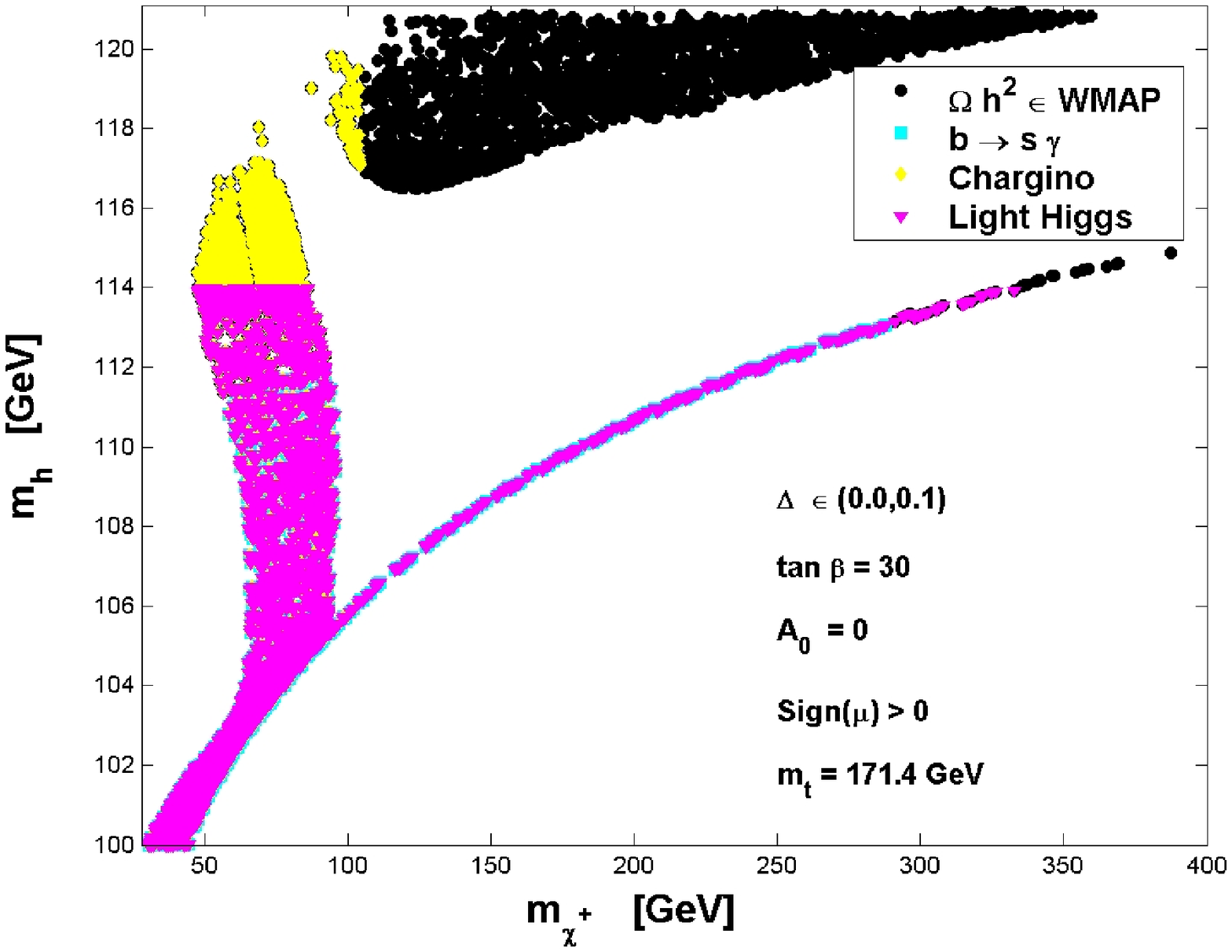}
\includegraphics[width=7.3cm,height=6.1cm]{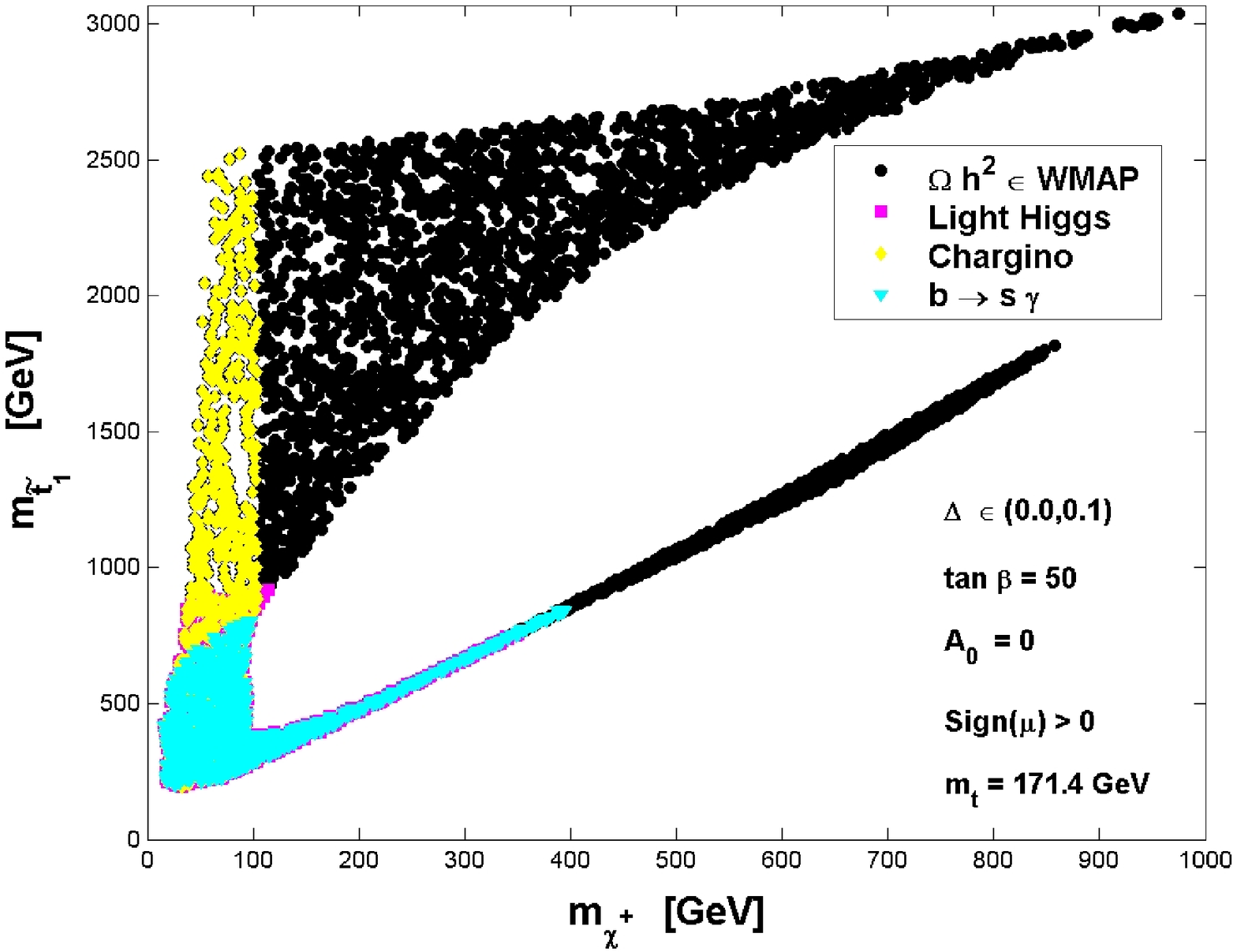}
\includegraphics[width=7.3cm,height=6.1cm]{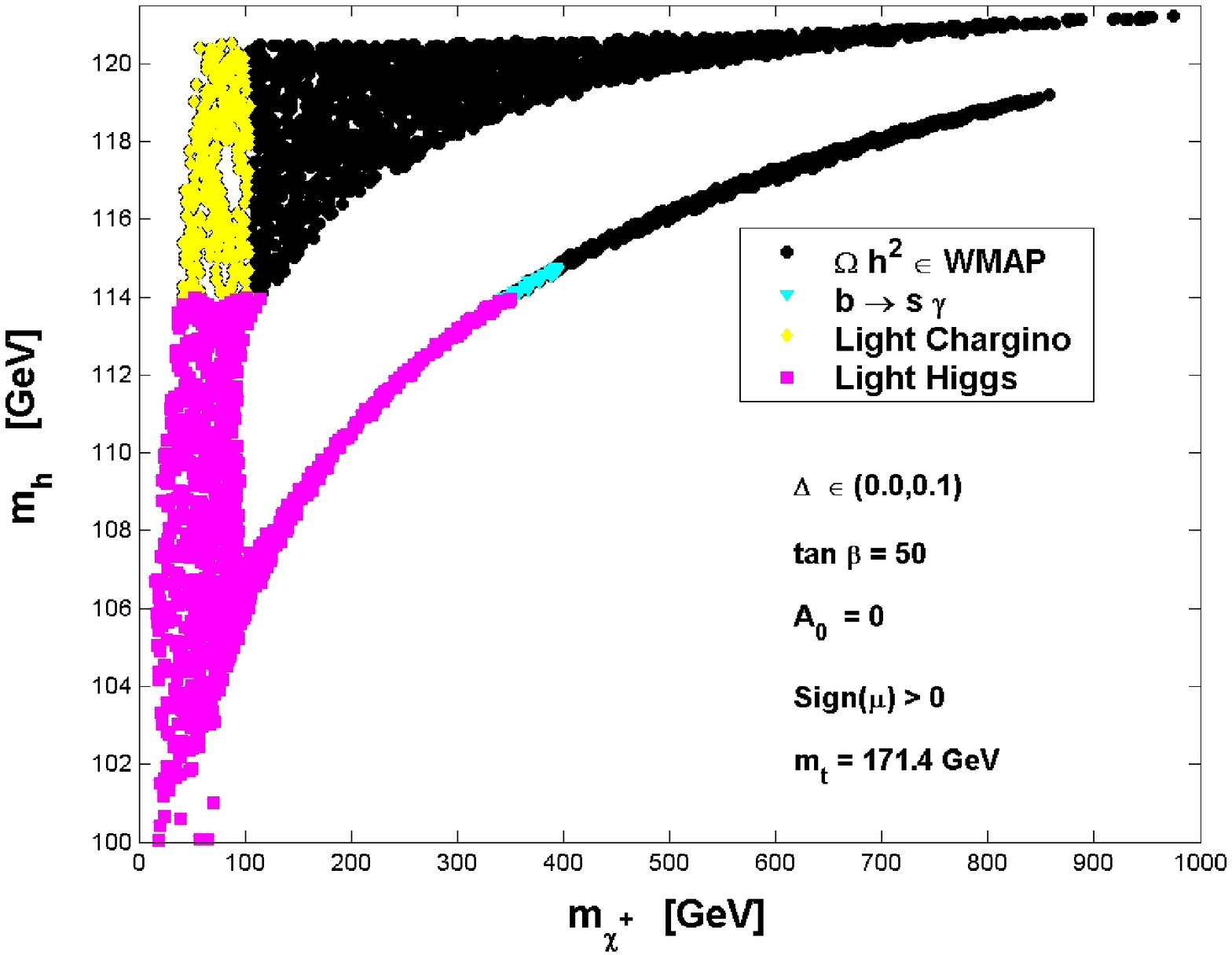}
\caption{ The allowed parameter space in the  $m_{\tilde{t}_1}
-m_{\chi^+}$ plane and in the   $m_h -m_{\chi^+}$   plane,  under
the $1\sigma$ WMAP3 constraint of  Eq.(\ref{wmap3}) in  extended
mSUGRA
 for the same data set as  in Fig.(\ref{fset1}).
 \label{fset4}}
\end{figure}

\clearpage

\begin{figure}[h]
\vspace*{1in} \hspace*{-.2in} \centering
\includegraphics[width=16cm,height=12cm]{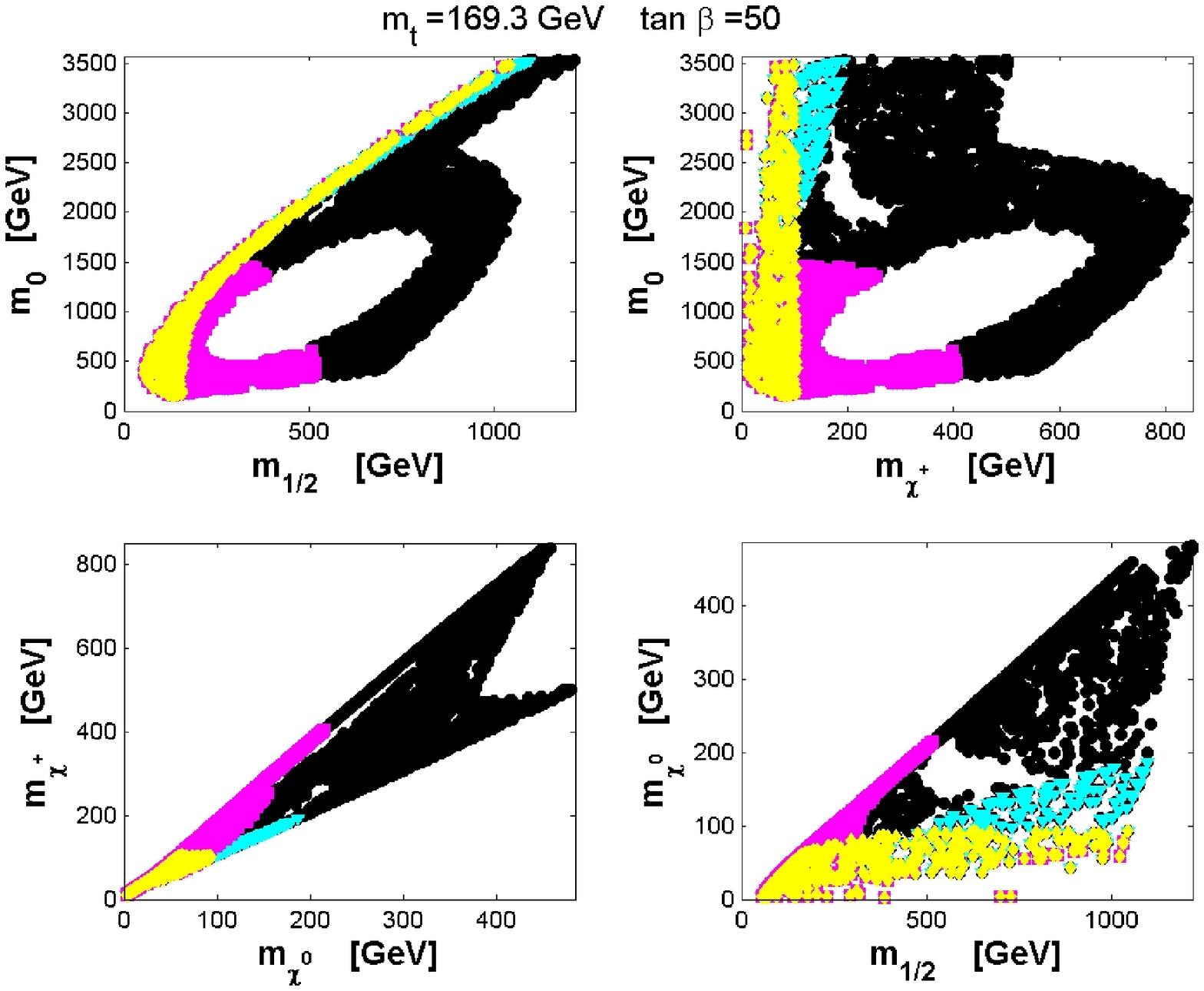}
\caption{An analysis for the case when $m_t=169.3$ GeV which is
$1\sigma$ below the central value, and $\tan\beta =50$ when all
other parameters  and constraints are  the same as  in
Fig.(\ref{fset1}). \label{fset5}}
\end{figure}

\begin{figure}[h]
\vspace*{1in} \hspace*{-.2in} \centering
\includegraphics[width=16cm,height=12cm]{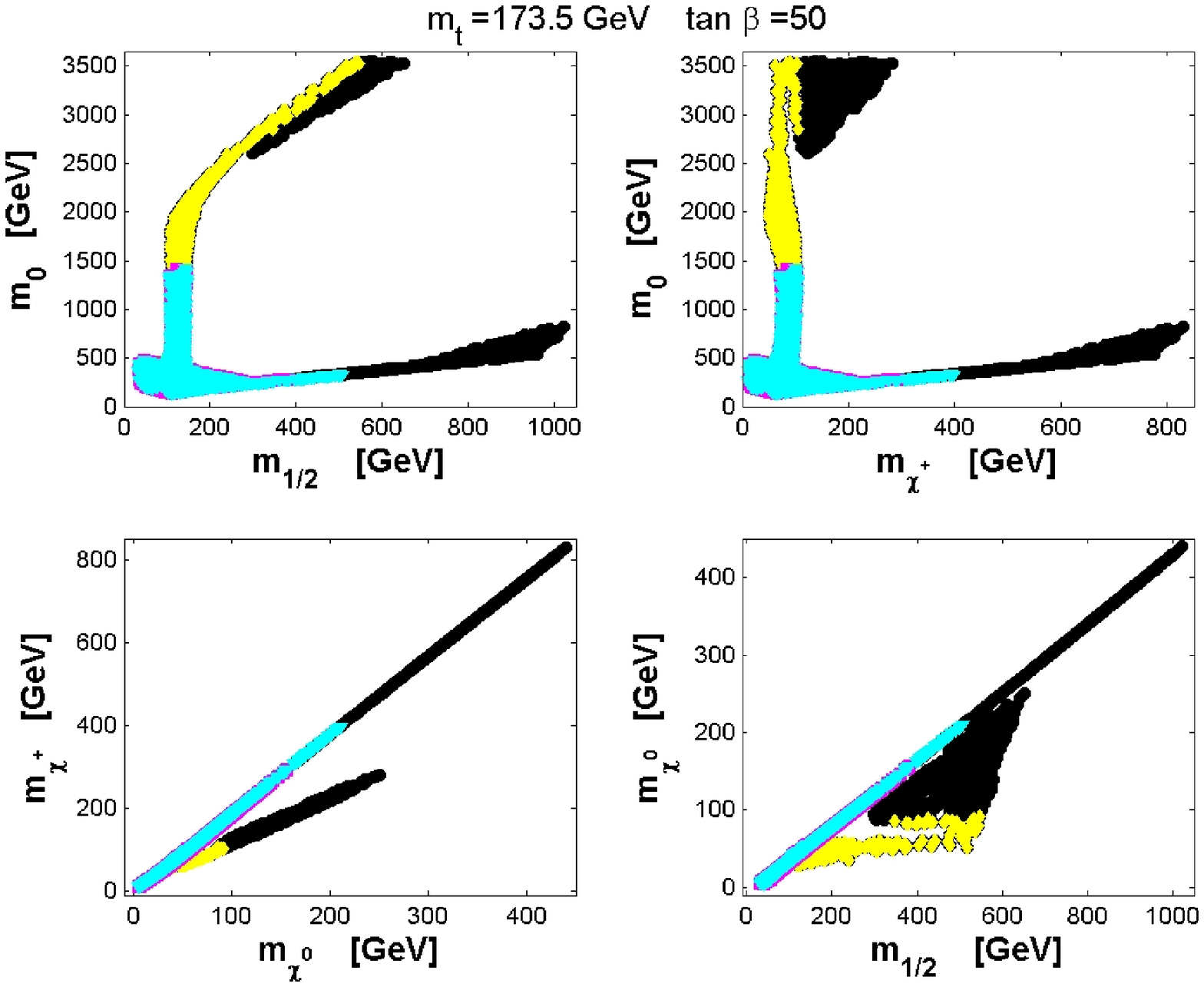}
\caption{An analysis for the case when $m_t=173.5$ GeV which is
$1\sigma$ above the central value, and $\tan\beta =50$ when all
other parameters  and constraints are  the same as  in
Fig.(\ref{fset1}). \label{fset6}}
\end{figure}

\end{document}